\documentclass{aa}

\usepackage[varg]{txfonts}

\defcitealias{Zoutendijk-2020-A&A-635-A107}{Paper~I}

\begin{document}
    \title{The MUSE-Faint survey}
    \subtitle{II.~The dark matter–density profile of the ultra-faint dwarf galaxy Eridanus~2\thanks{Based on observations made with ESO Telescopes at the La Silla Paranal Observatory under programme IDs 0100.D-0807, 0102.D-0372, 0103.D-0705, and 0104.D-0199.}}
    \titlerunning{The MUSE-Faint survey. II.}
    \author{%
        Sebastiaan~L. Zoutendijk\inst{1} \and
        Jarle Brinchmann\inst{2,1} \and
        Nicolas~F. Bouch\'e\inst{3} \and
        Mark den Brok\inst{4} \and
        Davor Krajnovi\'c\inst{4} \and
        Konrad Kuijken\inst{1} \and
        Michael V. Maseda\inst{1} \and
        Joop Schaye\inst{1}%
    }
    \authorrunning{S.~L.~Zoutendijk et al.}
    \institute{%
        Leiden Observatory, Leiden University, P.O.~Box~9513, 2300~RA~Leiden, The Netherlands\\\email{zoutendijk@strw.leidenuniv.nl} \and
        Instituto de Astrof\'{\i}sica e Ci\^encias do Espa\c{c}o, Universidade do Porto, CAUP, Rua das Estrelas, PT4150-762~Porto, Portugal \and
        Univ. Lyon, Univ. Lyon1, ENS de Lyon, CNRS, Centre de Recherche Astrophysique de Lyon UMR5574, 69230, Saint-Genis-Laval, France \and
        Leibniz-Institut f\"ur Astrophysik Potsdam (AIP), An der Sternwarte 16, D-14482 Potsdam, Germany%
    }
    \date{Received date / Accepted date}
    \abstract{%
    }{
        We use stellar line-of-sight velocities to constrain the dark matter--density profile of Eridanus~2, an ultra-faint dwarf galaxy with an absolute V-band magnitude $M_\mathrm{V} = -7.1$ corresponding to a stellar mass $M_* \approx 9 \times 10^4\,M_\sun$. We furthermore derive constraints on fundamental properties of self-interacting and fuzzy dark matter scenarios.%
    }{
        We present new observations of Eridanus~2 from MUSE-Faint, a survey of ultra-faint dwarf galaxies with the Multi Unit Spectroscopic Explorer on the Very Large Telescope, and determine line-of-sight velocities for stars inside the half-light radius.
        Combined with literature data, we have 92~stellar tracers out to twice the half-light radius.
        We constrain models of cold dark matter, self-interacting dark matter, and fuzzy dark matter with these tracers, using CJAM and pyGravSphere for the dynamical analysis.
        The models of self-interacting and fuzzy dark matter relate the self-interaction coefficient respectively the dark-matter particle mass to the density profile.%
    }{
        We find substantial evidence (Bayes factor ${\sim}10^{-0.6}$) for cold dark matter (a cuspy halo) over self-interacting dark matter (a cored halo) and weak evidence (Bayes factor ${\sim}10^{-0.4}$) for fuzzy dark matter over cold dark matter.
        We find a virial mass $M_{200} \sim 10^8\,M_\sun$ and astrophysical factors $J(\alpha_\mathrm{c}^J) \sim 10^{11}\,M_\sun^2\,\mathrm{kpc}^{-5}$ and $D(\alpha_\mathrm{c}^D) \sim 10^2$--$10^{2.5}\,M_\sun\,\mathrm{kpc}^{-2}$ (proportional to dark-matter annihilation and decay signals, respectively), the exact values depending on the density profile model.
        The mass-to-light ratio within the half-light radius is consistent with the literature.
        We do not resolve a core ($r_\mathrm{c} < 47\,\mathrm{pc}$, 68-\% confidence level) or soliton ($r_\mathrm{sol} < 7.2\,\mathrm{pc}$, 68-\% confidence level).
        These limits are equivalent to an effective self-interaction coefficient $f\Gamma < 2.2 \times 10^{-29}\,\mathrm{cm}^3\,\mathrm{s}^{-1}\,\mathrm{eV}^{-1}\,c^2$ and a fuzzy-dark-matter particle mass $m_\mathrm{a} > 4.0 \times 10^{-20}\,\mathrm{eV}\,c^{-2}$.
        The constraint on self-interaction is complementary to those from gamma-ray searches. The constraint on fuzzy-dark-matter particle mass is inconsistent with those obtained for larger dwarf galaxies, suggesting that the flattened density profiles of those galaxies are not caused by fuzzy dark matter.
    }{%
    }
    \maketitle

\section{Introduction}
\label{sec:introduction}
    Over time the astrophysical community has come to realize that baryonic matter and the established laws of physics are unable to explain our observations of the Universe.
    The discrepancy between baryonic and measured mass is almost universally interpreted as evidence for dark matter.
    The current paradigm, cold dark matter~(CDM), has so far been able to explain our observations, albeit with a few open questions.
    Various departures from the paradigm have been proposed with varying success, seeking to address a perceived shortcoming of CDM or to explain the properties of dark matter as a consequence of a more physically motivated theory.
    The proposed alternatives to CDM span a wide range of masses and interactions, including weakly interacting massive particles~(WIMPs; \citealt{Steigman-1985-NuPhB-253-375}), massive astrophysical compact halo objects~(MACHOs; \citealt{Griest-1991-ApJ-366-412}), axions~\citep{Weinberg-1978-PhRvL-40-223, Wilczek-1978-PhRvL-40-279, Preskill-1983-PhLB-120-127}, warm dark matter~(WDM) such as sterile neutrinos~\citep{Dodelson-1994-PhRvL-72-17}, and self-interacting dark matter~(SIDM; \citealt{Carlson-1992-ApJ-398-43, Spergel-2000-PhRvL-84-3760}).
    Another option, which has enjoyed less support, to solve the problem of `missing mass' is to modify the laws of gravity instead of adding extra mass to the Universe.
    Examples of these modifications are modified Newtonian dynamics~\citep{Milgrom-1983-ApJ-270-365} and emergent gravity~\citep{Verlinde-2017-ScPP-2-016}.
    In this paper we will limit ourselves to a few different forms of dark matter.

    The alternatives to CDM have different microphysical properties that lead to changes on astrophysical scales, making it in principle possible to distinguish between the individual alternatives and CDM through astronomical observations.
    One way of doing this is by investigating the gravitational interaction between the invisible dark matter and luminous objects.
    Different dark-matter theories often predict different spatial distributions of dark matter, which can be inferred from the kinematics of baryonic tracers.
    This kinematic approach is indirect, but is complementary to the direct and indirect approaches that search for signatures like annihilation and decay products.
    A complicating factor for astronomical observations is the complexity of astrophysical processes taking place in astronomical structures at the same time or in the past, which might also affect the measured spatial distribution of dark matter or the kinematics of the tracers.

    Ultra-faint dwarf galaxies~(UFDs) are perhaps the most promising class of objects for constraining dark matter on the basis of the density profile, because these galaxies are the most dark matter--dominated galaxies known \citep[see e.g.][]{McConnachie-2012-AJ-144-4} and also in an absolute sense contain very little baryonic matter that might interfere with the interpretation of the results ($M_\mathrm{V} > -7.7$; \citealt{Simon-2019-ARA&A-57-375}).
    Baryonic effects are expected to be able to create significant cores in larger dwarf galaxies~\citep{Brooks-2014-ApJ-786-87, DiCintio-2014-MNRAS-437-415}.
    Simulations of isolated galaxies show that the baryonic effects at play include bursty star formation, supernova feedback, and gas in- and outflows, or gravitational potential fluctuations in general~\citep[e.g.,][]{Read-2016-MNRAS-459-2573, ElZant-2016-MNRAS-461-1745, Freundlich-2020-MNRAS-491-4523}.
    Observational evidence that this process takes place in classical dwarf galaxies is found by \citet{Read-2019-MNRAS-484-1401}, who measure an anti-correlation between the dark-matter density at a radius of $150~\mathrm{pc}$ and the stellar-mass--to--halo-mass ratio.
    In the case of UFDs, the baryonic content is so low that it is not expected to significantly alter the density profile from cuspy to cored~\citep{Pennarrubia-2012-ApJ-759-L42, Onnorbe-2015-MNRAS-454-2092, Wheeler-2019-MNRAS-490-4447}.
    However, other effects such as tides~\citep{2020arXiv201109482G} can also create cores in a CDM universe, and non-circular motions can bias kinematic analyses to make cusps appear as cores~\citep{Oman-2019-MNRAS-482-821}.

    This paper is the second part of a series of papers on MUSE-Faint, a survey of ultra-faint dwarf galaxies with the Multi Unit Spectroscopic Explorer~(MUSE; \citealt{Bacon-2010-SPIE-7735-773508}) at the Very Large Telescope~(VLT).
    Previously~\citep[hereafter \citetalias{Zoutendijk-2020-A&A-635-A107}]{Zoutendijk-2020-A&A-635-A107} we have presented $4.5\,\mathrm{h}$ of observations on the central square arcminute of \object{Eridanus~2}~(Eri~2), a relatively bright UFD with absolute V-band magnitude $M_\mathrm{V} = -7.1$~\citep{Crnojevic-2016-ApJL-824-L14}.
    We found an intrinsic velocity dispersion of $10.3^{+3.9}_{-3.2}\,\mathrm{km}\,\mathrm{s}^{-1}$ for the bulk of the stars in the centre of Eri~2, whereas the central stellar overdensity was found to have an intrinsic velocity dispersion $<7.6\,\mathrm{km}\,\mathrm{s}^{-1}$ (68-\% confidence level), supporting the earlier photometric classification as a star cluster~\citep{Crnojevic-2016-ApJL-824-L14}.

    The kinematics of larger dwarf galaxies are well studied.
    \object{Fornax}, \object{Sculptor}, and \object{Draco}, for example, have large sets of stellar line-of-sight velocities~\citep{Walker-2009-AJ-137-3100, Walker-2015-MNRAS-448-2717} and the latter two even have internal proper motion measurements~\citep{Massari-2018-NatAs-2-156, Massari-2020-A&A-633-A36}.
    The profile of Fornax has been established as cored \citep[e.g.,][]{Goerdt-2006-MNRAS-368-1073, Walker-2011-ApJ-742-20, Amorisco-2013-MNRAS-429-L89}, whereas Draco is generally regarded as having a cuspy density profile \citep[e.g.,][]{Jardel-2013-ApJ-763-91, Read-2018-MNRAS-481-860, Massari-2020-A&A-633-A36}.
    There is no consensus about the density profile of Sculptor, with some authors preferring cores \citep[e.g.,][]{Battaglia-2008-ApJ-681-L13, Walker-2011-ApJ-742-20}, some cusps~\citep{Richardson-2014-MNRAS-441-1584, Massari-2018-NatAs-2-156}, and others claim either fits the data \citep[e.g.,][]{Breddels-2013-MNRAS-433-3173, Strigari-2018-ApJ-860-56}.
    However, \citet{Read-2019-MNRAS-484-1401} note that the enclosed mass estimates for Sculptor are in agreement, the largest tension being ${\sim}2\sigma$.

    Far fewer kinematic data are available for UFDs.
    The first UFD for which a velocity dispersion was determined was \object{Ursa Major~I}~\citep{Kleyna-2005-ApJ-630-L141}.
    Currently, velocity dispersions are known for over half of the confirmed and candidate UFDs~\citep{Simon-2019-ARA&A-57-375}.
    These measurements can be converted to mass estimates, for example using the estimators by \citet{Wolf-2010-MNRAS-406-1220}.
    Constraining a density profile for a UFD has so far not been possible due to the small sizes of the kinematic data sets and the limited radial ranges covered.
    However, the presence of the star cluster in Eri~2 has been used to argue for its hosting of a cored profile~\citep{Amorisco-2017-ApJ-844-64, Contenta-2018-MNRAS-476-3124}.

    Even without knowing the full density profile, classical and ultra-faint dwarf galaxies can be used to constrain dark-matter properties.
    If dark matter annihilates or decays, dark-matter haloes will emit radiation.
    Dwarf galaxies are promising targets because of their high dark-matter density and low radiation of baryonic origin.
    The annihilation and decay signals are proportional to the astrophysical $J$~and $D$~factors, which are integrated measures of the density profile.
    These factors are necessary to convert observed fluxes or flux limits to dark-matter properties.
    A number of studies have determined one or both of the astrophysical factors for dwarf galaxies \citep[e.g.,][]{Bonnivard-2015-MNRAS-453-849, FermiLAT-2014-PhRvD-89-042001, 2020arXiv200201229A}.

    Here we present additional observations from MUSE-Faint on four new pointings surrounding the centre, roughly covering the half-light radius of Eri~2, $R_{1/2}/D = 2.31\pm0.12\,\mathrm{arcmin}$ at distance $D = 366\pm17\,\mathrm{kpc}$, or $R_{1/2} = 277\pm14\,\mathrm{pc}$~\citep{Crnojevic-2016-ApJL-824-L14}.
    With these new fields, in combination with our central field and results from another study~\citep{Li-2017-ApJ-838-8} at larger distances from the centre, we can study the kinematics of stars in Eri~2 over a wide range of radii.
    Using different kinematical analysis techniques, we put constraints on the dark matter--density profile of Eri~2, specifically whether the profile is cuspy or cored and to what degree, and translate these to constraints on the properties of dark-matter candidates: the self-interaction coefficient of self-interacting dark matter and the dark-matter particle mass of fuzzy dark matter.
    Furthermore, we compare different models to each other using the Bayesian evidence, in an attempt to constrain which kinds of dark matter fit the data better.
    In the figures in this paper we consistently assign a colour to each model of the density profile to facilitate recognition and association.

    In Sect.~\ref{sec:methods} we describe our data and its reduction (Sect.~\ref{ssec:obsred}), the dark matter--density profile models (Sect.~\ref{ssec:models}), and the analysis methods (Sects.~\ref{ssec:cjam} and~\ref{ssec:pgsph}).
    We continue in Sect.~\ref{sec:results} with our results on dark-matter parameter constraints (Sect.~\ref{ssec:param}), density-profile recovery and derived halo properties (Sect.~\ref{ssec:recovery}), and a comparison of the evidence for the different dark-matter models (Sect.~\ref{ssec:modcomp}).
    We end with a discussion in Sect.~\ref{sec:discussion} and our conclusions in Sect.~\ref{sec:conclusions}.

\section{Methods}
\label{sec:methods}
    We begin by describing our observations of Eri~2 from the MUSE-Faint survey, the data reduction, and the extraction of kinematics in Sect.~\ref{ssec:obsred}.
    This is followed in Sect.~\ref{ssec:models} by the presentation of the three main dark-matter models tested in this paper.
    The parameters of the density profiles associated with these models are linked to microphysical properties of dark matter.
    To constrain the profiles and thereby these properties, we use two analysis tools, CJAM and pyGravSphere, introduced in Sects.~\ref{ssec:cjam} and~\ref{ssec:pgsph}.

\subsection{Observations and data reduction}
\label{ssec:obsred}
    % 2017-10-16 GTO-20-P100 0100.D-0807 0.72-0.97 0.7-1.0
    % 2017-10-18 GTO-20-P100 0100.D-0807 0.72-0.76 0.6
    % 2017-10-19 GTO-20-P100 0100.D-0807 0.67-0.74 0.5-0.6
    % 2017-10-20 GTO-20-P100 0100.D-0807 0.68-0.74 0.5-0.6
    % 2017-10-21 GTO-20-P100 0100.D-0807 0.77-0.80 SGS0.40-0.60
    % 2017-10-24 GTO-20-P100 0100.D-0807 0.62-0.65 SGS0.37-0.71
    % 2017-11-16 GTO-21-P100 0100.D-0807 0.60-0.69 ~0.6
    % 2018-02-14 GTO-22-P100 0100.D-0807 0.73-0.93 0.8-1.2
    % 2018-02-15 GTO-22-P100 0100.D-0807 0.88-0.92 0.9-1.0
    % 2018-03-15 GTO-23-P100 0100.D-0807 0.83-0.84 ~0.8
    % 2018-03-16 GTO-23-P100 0100.D-0807 0.80-0.86 ~0.8
    % 2019-03-04 GTO-32-P102 0102.D-0372 0.72-0.80 0.7
    % 2019-09-27 GTO-35-P103 0103.D-0705 0.68-0.70 SGS0.43-0.52
    % 2019-09-29 GTO-35-P103 0103.D-0705 0.87-1.02 0.8-1.5
    % 2019-10-28 GTO-36-P104 0104.D-0199 0.66-0.72 0.5-0.6
    % 2019-10-29 GTO-36-P104 0104.D-0199 0.74-0.84 0.5-0.6
    % 2019-11-23 GTO-37-P104 0104.D-0199 0.81-0.85 0.6
    % 2019-11-26 GTO-37-P104 0104.D-0199 0.67-0.80 0.49
    % 2019-12-24 GTO-38-P104 0104.D-0199 0.73-0.75 SGS0.36-0.68
    % 2019-12-25 GTO-38-P104 0104.D-0199 0.71-0.80 SGS0.45-0.57

\begin{figure}
        \includegraphics[width=\linewidth]{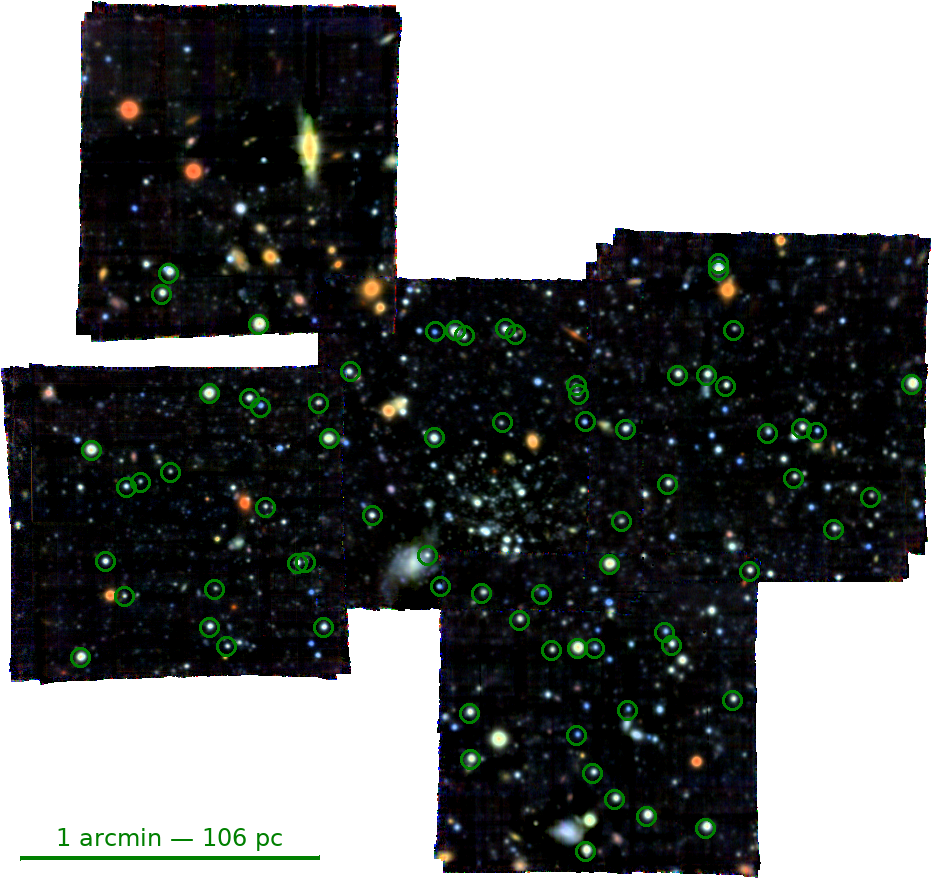}
        \caption{%
            Composite-colour image mosaic of Eridanus~2 as observed with MUSE-Faint.
            Sloan Digital Sky Survey filters g, r, and i were used for the colours blue, green, and red, respectively.
            Images of the five separately reduced fields were combined with Montage and the colours were composited using the algorithm of \citet{Lupton-2004-PASP-116-133}.
            The 72 member stars with MUSE-Faint measurements are circled in green.
            Celestial north is up.
            The angular and physical scale at the distance of Eridanus~2 are indicated in the bottom left corner.%
        }
        \label{fig:im}
    \end{figure}

    The data were taken with VLT/MUSE during five guaranteed-time observing runs between October 2017 and December 2019.
    The estimated natural seeing varied between $0.6$ and $1.0\,\mathrm{arcsec}$, with adaptive optics reducing the seeing by $0.1$--$0.2\,\mathrm{arcsec}$ under good conditions.
    In \citetalias{Zoutendijk-2020-A&A-635-A107} we described the data reduction and source selection for Field~1, our central pointing on Eri~2.
    We used the same procedure independently on Fields~2 through~5, presented here for the first time (see Fig~\ref{fig:im}).
    In brief, we mostly followed the standard procedure of reducing MUSE data with the MUSE Data Reduction Software (DRS; version~2.4 for Field~1 and version~2.6 for Fields~2 through~5; \citealt{2020arXiv200608638W}), the exceptions being the use of the bad-pixel table from \citet{Bacon-2017-A&A-608-A1} and an autocalibration step on a source-masked version of the cube.
    The DRS-produced data cubes were post-processed with the Zurich Atmosphere Purge~(ZAP; version~2.0; \citealt{Soto-2016-MNRAS-458-3210}) to remove residual sky signatures.
    We extracted spectra from these data cubes using PampelMuse~\citep{Kamann-2013-A&A-549-A71} and measured seeing full widths at half-maximum between~$0.53$ and~$0.66\,\mathrm{arcsec}$ at $7000\,\mathrm{\AA}$ for the five data cubes, using public \emph{Hubble Space Telescope} data\footnote{\emph{Hubble Space Telescope} proposal GO-14234, principal investigator J.\,D.~Simon, presented by \citet{2020arXiv201200043S}.} to construct a source catalogue.
    We used spexxy~(version~2.5; \citealt{Husser-2012-3DSDSP-UG-0}) with the PHOENIX library of synthetic stellar spectra to determine line-of-sight velocities and made a catalogue of the results for each field.
    To ensure reliable velocity measurements and to limit contamination from background galaxies and Milky-Way stars, we imposed a set of selection criteria: we removed catalogue entries that have a clearly extra-galactic spatial or spectral appearance, a spectral signal-to-noise ratio below 5, an unsuccessful velocity fit, a parallax measurement from \emph{Gaia} Data Release~2~\citep{GaiaCollaboration-2016-A&A-595-A1, GaiaCollaboration-2018-A&A-616-A1, Lindegren-2018-A&A-616-A2} inconsistent with zero, and photometry inconsistent with a broadened MIST isochrone~\citep{Dotter-2016-ApJS-222-8, Choi-2016-ApJ-823-102, Paxton-2011-ApJS-192-3, Paxton-2013-ApJS-208-4, Paxton-2015-ApJS-220-15}.
    We had 95 entries that passed these criteria in the five catalogues.
    To this we added another catalogue with 47~observations of 28~member stars identified by \citet{Li-2017-ApJ-838-8}, bringing the total number of entries to 142.

    Since the six catalogues have some overlap on the sky, some sources occur in multiple catalogues.
    While merging the six source catalogues, we took into account the presence of these duplicate entries, which share their identifier, by replacing them with a single entry in the final catalogue, where we took the mean values of the right ascensions and declinations, the uncertainty-weighted mean values of the line-of-sight velocities, the sum in quadrature of the inverse uncertainties on the line-of-sight velocities, and the sum in quadrature of the signal-to-noise ratios.
    After this removal of duplicates, we were left with 109~unique stars.
    As in \citetalias{Zoutendijk-2020-A&A-635-A107}, we checked for possible remaining contamination of our sample by Milky-Way stars by computing the membership probabilities of the selected sources.
    This we did by calculating the likelihood of observing the measured stellar velocities given two distribution functions, a Gaussian representing Eri~2 and a contaminating distribution based on the Besan\c{c}on model of the Milky Way~\citep{Robin-2003-A&A-409-523, Robin-2004-A&A-416-157}, and a membership probability for each star weighting the contributions of both distribution functions.
    The membership probabilities were determined by optimizing the likelihood while marginalizing over the mean velocity and dispersion of Eri~2.
    We found that 10 of our sources have significantly lower membership probabilities than the others, leading to their exclusion from our sample, leaving 99~stars.

    In \citetalias{Zoutendijk-2020-A&A-635-A107} we found that the Eri~2 cluster seen at the centre of this galaxy has a different kinematic distribution than the bulk of Eri~2.
    Moreover, it is still not completely clear how far this cluster is located from the centre of Eri~2, as we can only see the projected location.
    This leads to the question of whether the kinematics of the stars that make up the cluster are good tracers of the potential of Eri~2, or whether they trace mainly the properties of the star cluster itself.
    To avoid a possible bias in our results, we excluded the seven cluster member stars identified in \citetalias{Zoutendijk-2020-A&A-635-A107} from our sample, bringing our final selection to 92~stars.
    We present the positions and kinematics of the final selection in Table~\ref{tab:selection}.
    Of the final selection, 64~stars have only MUSE-Faint measurements, 20~stars have only measurements from \citet{Li-2017-ApJ-838-8}, and eight stars have measurements from both sources.

\subsection{Models of dark matter--density profiles}
\label{ssec:models}
    With the goal to place constraints on the nature of dark matter, we will compare our kinematic data to several models of dark matter--density profiles, each based on a different type of dark matter.
    As a null hypothesis, we will use a Navarro--Frenk--White~(NFW; \citealt{Navarro-1996-ApJ-462-563}) profile to represent cold dark matter~(CDM):
    \begin{equation}
        \rho_\mathrm{CDM}(r; \rho_0, r_\mathrm{s}) = \frac{\rho_0}{(r/r_\mathrm{s})(1 + r/r_\mathrm{s})^2},
        \label{eq:cdm}
    \end{equation}
    where $\rho_{0}$ is known as the characteristic density and $r_\mathrm{s}$ is the scale radius.
    We will compare this with two other models: self-interacting dark matter~(SIDM) and fuzzy dark matter~(FDM).
    The latter two models behave like an NFW profile on large scales, but deviate on smaller scales, depending on the effective self-interaction coefficient and the  mass of the dark-matter particle, respectively.
    We can therefore, for both the SIDM and the FDM model, not only compare one dark-matter theory to the other, but also place constraints on the properties of dark-matter particles under the assumption of the particular theory.

    SIDM describes a form of dark matter that interacts with itself more strongly than with other particles~\citep{Spergel-2000-PhRvL-84-3760}.
    Interactions that remove dark-matter particles from the halo according to the relation
    \begin{equation}
        \dot{\rho}(x, t) = -\Gamma\rho^2(x, t),
        \label{eq:selfint}
    \end{equation}
    where $\Gamma$ is the self-interaction coefficient, produce a density profile
    \begin{equation}
        \rho_\mathrm{SIDM}(r; \rho_\mathrm{c}, r_\mathrm{c}, r_\mathrm{s}) = \frac{\rho_\mathrm{c}}{(r/r_\mathrm{c})(1 + r/r_\mathrm{s})^2 + 1},
        \label{eq:sidm1}
    \end{equation}
    where $\rho_\mathrm{c}$ is the core density and $r_\mathrm{c}$ is the core radius~\citep{Lin-2016-JCAP-03-009}.
    We will discuss how $\Gamma$ and our constraints thereon relate to the cross section~$\sigma$ in Section~\ref{sec:discussion}.
    The self-interaction described covers scattering and annihilation, but has been designed with mainly the latter in mind.
    The profile can also be written as
    \begin{equation}
        \rho_\mathrm{SIDM}(r; \rho_0, r_\mathrm{c}, r_\mathrm{s}) = \frac{\rho_0}{r_\mathrm{c}/r_\mathrm{s} + (r/r_\mathrm{s})(1 + r/r_\mathrm{s})^2}
        \label{eq:sidm2}
    \end{equation}
    with characteristic density $\rho_0 = \rho_\mathrm{c}(r_\mathrm{c}/r_\mathrm{s})$.
    The SIDM profile is equal to the CDM/NFW profile for $r_\mathrm{c} = 0$, but for $r_\mathrm{c} > 0$ it exhibits a core instead of a cusp.
    Evidence in favour of the SIDM profile over the CDM profile would indicate that the density profile of Eri~2 is cored.
    If the density profile of Eri~2 is cuspy, both the CDM and SIDM model should be able to describe it, but we should in this case find evidence in favour of the CDM profile, as it is the simpler of the two.
    At large radii the SIDM profile always asymptotes to the NFW profile.
    There is a relation tying the self-interaction coefficient~$\Gamma$ of the dark matter to the observational properties of the profile~\citep{Lin-2016-JCAP-03-009}:
    \begin{equation}
        f\Gamma = \frac{r_\mathrm{c}/r_\mathrm{s}}{t \rho_0},
        \label{eq:fGamma}
    \end{equation}
    where $t$ is the time elapsed since the start of the self-interaction, at the virialization of the dark-matter halo.
    However, this relation is degenerate with the fudge factor~$f$ that compensates for the unknown gravitational back-reaction.
    As dark-matter particles interact according to Equation~\eqref{eq:selfint}, the dark-matter halo moves out of dynamical equilibrium.
    The gravitational back-reaction is the process of the halo re-adjusting to the new dynamical equilibrium, thereby altering the profile to a larger extent than described by $\Gamma$ alone.
    The value of~$f$ is estimated to be ${\sim}10$ for dwarf galaxies~\citep{Kaplinghat-2000-PhRvL-85-3335}, but is not precisely known.
    We will therefore try to constrain the product $f\Gamma$, which we will call the effective self-interaction coefficient.
    The time~$t$ is not known, so we assume it is equal to the age of the stellar population.
    This was estimated to be $8\,\mathrm{Gyr}$ in \citetalias{Zoutendijk-2020-A&A-635-A107}, but in a more rigorous analysis \citet{2020arXiv201200043S} find the oldest stars are ${\sim}13.5\,\mathrm{Gyr}$ old, therefore we adopt the latter.
    Should a better estimate of the time since virialization become available in the future, our constraints of $f\Gamma$ can simply be rescaled.

    FDM consists of ultra-light spinless bosons that form a Bose--Einstein condensate, exhibiting quantum-mechanical behaviour at astronomical scales~\citep{Hu-2000-PhRvL-85-1158}.
    Axions are a possible and well-motivated class of particles that can form FDM, but are not the only possibility, nor does FDM require an electromagnetic interaction like axions have~\citep[see e.g.][]{2020arXiv200503254F}.
    The wave-like properties of FDM result in a density profile~\citep{Schive-2014-NatPh-10-496, Schive-2014-PhRvL-113-261302, Marsh-2015-MNRAS-451-2479}
    \begin{equation}
        \rho_\mathrm{FDM}(r; \rho_{\mathrm{sol},0}, r_\mathrm{sol}, \rho_{\mathrm{CDM},0}, r_\mathrm{s}) =
        \begin{cases}
            \rho_\mathrm{sol}(r; \rho_{\mathrm{sol},0}, r_\mathrm{sol}), & (r < r_\mathrm{t}),\\
            \rho_\mathrm{CDM}(r; \rho_{\mathrm{CDM},0}, r_\mathrm{s}), & (r \geq r_\mathrm{t}),\\
        \end{cases}
        \label{eq:fdm}
    \end{equation}
    where
    \begin{equation}
        \rho_\mathrm{sol}(r; \rho_{\mathrm{sol},0}, r_\mathrm{sol}) = \frac{\rho_{\mathrm{sol},0}}{(1 + (r/r_\mathrm{sol})^2)^8}.
        \label{eq:sol}
    \end{equation}
    At large radii FDM follows the NFW profile, but with decreasing radius the density first rises steeply and then flattens to a constant value.
    This inner part of the profile deviating from the NFW is known as the soliton solution to the wave equations governing the ultra-light dark-matter particles, with central density~$\rho_{\mathrm{sol},0}$ and soliton radius $r_\mathrm{sol}$.
    We note that this soliton radius $r_\mathrm{sol}$, defined by \citet{Marsh-2015-MNRAS-451-2479}, differs from the soliton radius~$r_\mathrm{c}$ as defined by \citet{Schive-2014-NatPh-10-496}.
    The central soliton density and soliton radius are related to the mass of the dark-matter particle through
    \begin{equation}
        m_\mathrm{a} = \sqrt{\frac{2 \hbar M_\mathrm{Pl}^2}{\alpha^4 c r_\mathrm{sol}^4 \rho_{\mathrm{sol},0}}},
        \label{eq:ma}
    \end{equation}
    where $M_\mathrm{Pl}$ is the reduced Planck mass, $\alpha \approx 0.230$, and $c$ is the speed of light~\citep{Marsh-2015-MNRAS-451-2479}.
    There is a sharp transition, at the transition radius~$r_\mathrm{t}$, to an NFW profile.
    The profile has to be continuous (i.e., the two parts need to be equal at the transition radius), but the transition is so sharp that it is usually modelled with a sudden transition, leading to a discontinuous first derivative.
    Our method, however, necessitates a smooth modelling of the transition and is introduced in Sect.~\ref{ssec:cjam} and detailed in Appendix~\ref{app:mge}.
    The transition radius can be expressed in terms of the fraction~$\varepsilon$ of the density at the transition relative to the central soliton density~$\rho_{\mathrm{sol},0}$:
    \begin{equation}
        r_\mathrm{t} = (\varepsilon^{-1/8} - 1)^{1/2} r_\mathrm{sol}.
        \label{eq:rt}
    \end{equation}
    Simulations show that $\varepsilon$ does not exceed $1/2$~\citep{Schive-2014-NatPh-10-496, Marsh-2015-MNRAS-451-2479}.

    To be able to test the different dark matter--density profiles against our data, we need to make predictions for measurements given a set of parameters.
    This is not an easy task, considering that we only measure the projected positions of stars and their line-of-sight velocities.
    Converting between the three-dimensional models and the two-dimensional measurements leads to a dependence on the velocity anisotropy.
    This has long been a source of uncertainty for density profile determination, because it leads to a mass--anisotropy degeneracy when the enclosed mass is determined from the three-dimensional velocity dispersion through Jeans analysis.
    Fortunately, several methods exist that attempt to break this degeneracy by exploiting additional information available in the data.
    We will use two different codes in this paper, which take different approaches to the problem, each with its own merits and shortcomings.

\subsection{CJAM}
\label{ssec:cjam}
    The light and dark matter distributions can be approximated with a multi-Gaussian expansion~(MGE; \citealt{Emsellem-1994-A&A-285-723}).
    This approximation makes it possible to calculate integrals over the profiles analytically instead of numerically and leads to faster performance.
    The first method, CJAM~\citep{Watkins-2013-MNRAS-436-2598} is an implementation of the Jeans Anisotropic MGE method~(JAM; \citealt{Cappellari-2008-MNRAS-390-71}).
    CJAM calculates the first and second moments of the velocities for every tracer, allowing for non-spherical light and matter distributions and a non-zero, constant velocity anisotropy.
    In general, the first moments form a three-dimensional expectation value of the velocity of a tracer given a model and the nine second moments make up the covariance.
    As we only have line-of-sight information, we are limited to the first and second moments along the line of sight, though CJAM can also calculate moments in the plane of the sky, which could be compared to proper-motion data.
    Because of the limited number of available tracers, we also assume the dark-matter component of Eri~2 is spherically symmetric.
    The use of MGEs in CJAM allows us to implement our own density profiles.
    We describe the expansion of our profiles into MGEs in Appendix~\ref{app:mge}.

    There are several parametrizations in which we can express the different dark-matter profiles.
    We define the astrophysical parametrizations as those using astrophysical measurements such as characteristics densities and scale radii.
    These are the same as the canonical forms of the profiles, as given in Equations~\eqref{eq:cdm}--\eqref{eq:sol}.
    For SIDM and FDM we can transform the astrophysical parametrization into a microphysical parametrization.
    These parametrizations contain parameters that characterize dark-matter physics: the effective self-interaction coefficient and the dark matter--particle mass.
    However, we find that we get the best constraints by parametrizing the profiles using quantities that are as close as possible to our measurements.
    We will refer to these last parametrizations as computational.
    We constrain the computational parametrizations directly and compute the constraints on the astrophysical and microphysical parametrizations from them.

    For the SIDM profile, we found a computational parametrization in terms of the base-10 logarithm of dark-matter density at three fixed radii: $\log_{10} \rho_1$ at $r_1 = 50\,\mathrm{pc}$, $\log_{10} \rho_2$ at $r_2 = 100\,\mathrm{pc}$, and $\log_{10} \rho_3$ at $r_3 = 150\,\mathrm{pc}$.
    These radii are chosen to be near the peak in observed line-of-sight velocities.
    The astrophysical parameters can be recovered through
    \begin{align}
        r_\mathrm{s} &= r_1 \cdot \frac{(\rho_1-\rho_2)(9\rho_3-\rho_1)-(\rho_1-\rho_3)(4\rho_2-\rho_1)}{(\rho_1-\rho_2)(\rho_1-3\rho_3)-(\rho_1-\rho_3)(\rho_1-2\rho_2)},
        \label{eq:param1}\\
        r_\mathrm{c} &= r_1 \cdot \frac{(4\rho_2-\rho_1)(r_1/r_\mathrm{s}) - (\rho_1-2\rho_2)}{\rho_1-\rho_2}.
        \label{eq:param2}
    \end{align}
    As a special case with $r_\mathrm{c} = 0$, the CDM profile needs only two parameters, which simplifies the system of equations, yielding the solution
    \begin{equation}
        r_\mathrm{s} = r_1 \cdot \frac{9\rho_3-4\rho_2}{2\rho_2-3\rho_3}.
        \label{eq:param3}
    \end{equation}

    The consequence of this choice of parametrization is that it is harder to set a prior that will limit the astrophysical parameters to reasonable values.
    One could try to find a prior volume on the computational parameters that translates to the desired prior volume on the astrophysical parameters, but given the complexity of Equations~\eqref{eq:param1}--\eqref{eq:param3}, this is difficult and would introduce a non-trivial prior distribution.
    Instead, we choose to simply reject the points that translate to values outside the desired astrophysical priors by assigning them a probability of zero.
    We accept combinations of parameters that lead to values of $r_\mathrm{s}$ and $r_\mathrm{c}$ such that $10^{-2}\,r_\mathrm{s} \leq r_\mathrm{c} \leq r_\mathrm{s}$ and $10^{-3}\,r_\mathrm{s} \leq R_i \leq 10^3\,r_\mathrm{s}$, where $R_i$ is the projected radius of a tracer, for all tracers.
    These ranges are those over which the MGEs were fitted and should be sufficiently large to encompass all reasonable models for Eri~2.
    These cuts of unphysical and unreasonable parameter combinations are performed after sampling from the prior distribution, during the evaluation of the likelihood function.

    For the FDM profile, which is more complex due to the variable transition radius between the two different regimes, we were not able to find a similar parametrization in densities only.
    We therefore use a computational parametrization in the following parameters: the logarithm $\log_{10} \rho_{\mathrm{CDM},100} \coloneqq \log_{10} \rho_\mathrm{CDM}(100\,\mathrm{pc})$ of the outer density profile at $100\,\mathrm{pc}$, the logarithmic slope $\alpha_{\mathrm{CDM},100} \coloneqq (\mathrm{d}\ln\rho_\mathrm{CDM}/\mathrm{d}\ln r)(100\,\mathrm{pc})$ of the outer density profile at $100\,\mathrm{pc}$, the logarithm $\log_{10} (r_\mathrm{sol}/r_\mathrm{s})$ of the ratio between the soliton radius and scale radius, and the logarithm $\log_{10} \varepsilon = \log_{10} \rho_\mathrm{FDM}(r_\mathrm{t}) - \log_{10} \rho_{\mathrm{sol},0}$ of the density at the transition radius relative to the soliton density.

    We use MultiNest~\citep{Feroz-2008-MNRAS-384-449, Feroz-2009-MNRAS-398-1601, Feroz-2019-OJAp-2-10} through the PyMultiNest interface~\citep{Buchner-2014-A&A-564-A125} to find the posterior likelihood distribution for the parameters of each model, which consist of the aforementioned profile parameters and the systemic velocity~$v_0$ against which the kinematics are offset, using uniform priors over large ranges of values, listed in Table~\ref{tab:cjampri}.
    \begin{table}
        \caption{Limits of the uniform CJAM/MultiNest priors and to which profiles they apply.}
        \label{tab:cjampri}
        \centering
        \begin{tabular}{lccc}
            \hline
            \hline
            Prior & Min. & Max. & Profiles \\
            \hline
            $\log_{10}(\rho_1/M_\sun\,\mathrm{kpc}^{-3})$\tablefootmark{(a)} & $6$ & $12$ & SI \\
            $\log_{10}(\rho_2/M_\sun\,\mathrm{kpc}^{-3})$\tablefootmark{(a)} & $6$ & $12$ & C, SI \\
            $\log_{10}(\rho_3/M_\sun\,\mathrm{kpc}^{-3})$\tablefootmark{(a)} & $6$ & $12$ & C, SI \\
            $\log_{10}(\rho_{\mathrm{CDM},100}/M_\sun\,\mathrm{kpc}^{-3})$ & $6$ & $10$ & F \\
            $\alpha_{\mathrm{CDM},100}$ & $-3$ & $-1$ & F \\
            $\log_{10}(r_\mathrm{sol}/r_\mathrm{s})$ & $-3$ & $0$ & F \\
            $\log_{10} \varepsilon$ & $-5$ & $\log_{10} 1/2$ & F\\
            $v_0/\mathrm{km}\,\mathrm{s}^{-1}$ & 65 & 85 & C, SI, F \\
            \hline
        \end{tabular}
        \tablefoot{%
            The letters C, SI, and F indicate cold dark matter~(CDM), self-interacting dark matter~(SIDM), and fuzzy dark matter~(FDM), respectively.
            The parameters are: the densities $\rho_1$, $\rho_2$, and $\rho_3$ at $50$, $100$, and $150\,\mathrm{pc}$, respectively; the density~$\rho_{\mathrm{CDM},100}$ of the outer FDM profile at $100\,\mathrm{pc}$; the logarithmic slope~$\alpha_{\mathrm{CDM},100}$ of the outer FDM profile at $100\,\mathrm{pc}$; the ratio~$r_\mathrm{sol}/r_\mathrm{s}$ of the soliton radius to the scale radius; the relative density~$\varepsilon$ with respect to the central density at the transition radius between the inner and outer FDM profiles; and the systemic velocity~$v_0$.
            The parameter spaces of the CDM and SIDM models contain combinations of parameters that translate to unreasonable values for $r_\mathrm{c}$ and $r_\mathrm{s}$.
            This is handled by setting the likelihood in these regions to zero, but can also be thought of as being excluded from the prior space indicated.
            \tablefoottext{a}{Within the indicated priors, $\rho_i \geq \rho_{i+1}$.}%
        }
    \end{table}
    MultiNest also calculates the Bayesian evidence for each model, allowing us to compare the models with each other.
    The wide priors do not significantly impact the Bayesian evidence calculation because they extend to regions of parameter space with very low likelihoods.
    Since we exclude some models from consideration, one might be concerned that this compromises the Bayesian evidence calculation of MultiNest.
    We performed a few mock runs of MultiNest with a simple likelihood function to test whether our forcing of likelihoods to zero would affect the evidence calculation, as opposed to limiting the prior volume.
    We found that some of the evidence estimators are indeed biased, but not the nested sampling global log-evidence.
    We will therefore use this estimator to evaluate the Bayesian evidence of the models.

\subsection{pyGravSphere}
\label{ssec:pgsph}
    The second method we use to determine density profiles is GravSphere~\citep{Read-2017-MNRAS-471-4541}.
    Like the classical Jeans analysis, the GravSphere method directly calculates the dispersion of the measured line-of-sight velocities in bins at different radii, as opposed to the non-binned treatment of velocity expectation values done in JAM.
    What GravSphere adds, is that it can work with non-constant velocity anisotropies and that it calculates two higher-order moments in the radial bins, the virial shape parameters~(VSPs; \citealt{Merrifield-1990-AJ-99-1548}).
    These should partially break the degeneracy between mass and anisotropy that is present when only using the dispersion.
    A drawback is that GravSphere only allows for spherical symmetry, whereas JAM can handle axisymmetric distributions.

    We use the pyGravSphere implementation~\citep{2020MNRAS.498..144G} of the GravSphere method.
    We provide it with the same kinematic information as CJAM.
    To determine the tracer profile, we make a mock photometric catalogue drawing stars from the same exponential distribution as assumed for CJAM.
    We modify pyGravSphere to make the bin size configurable and to add remaining sources to the last (outer) bin.
    We divide the 92~sources with line-of-sight velocities into bins of 11, making eight bins, with four extra stars in the last bin.
    We also implement new estimators of the velocity moments and their uncertainties, designed to minimize the biases present in cases with large measurement uncertainties and few data.
    These unbiased estimators and their derivation are introduced in Appendix~\ref{app:vmom}.
    The estimators return a negative result for the velocity dispersion in bins 3 and 6.
    These bins are therefore discarded by pyGravSphere, leaving six bins in the analysis.
    We do not use the VSPs because there are too few stars per bin to accurately estimate their uncertainties.
    We explain this in more detail in Appendix~\ref{app:vmom}.
    Lastly, we modify pyGravSphere to place the estimators at the average projected radius of the stars in the corresponding bins, instead of at the maximum radius.
    The modified pyGravSphere binning code is made publicly available\footnote{\url{https://github.com/slzoutendijk/hkbin}} as a stand-alone program called hkbin.
    We show the binned data that pyGravSphere uses in Table~\ref{tab:pgsdata}.
    \begin{table}
        \caption{%
            Kinematic data of Eridanus~2 after binning, as used by pyGravSphere.%
        }
        \label{tab:pgsdata}
        \centering
        \begin{tabular}{cc}
            \hline
            \hline
            Radius ($\mathrm{kpc}$) & Velocity dispersion ($\mathrm{km}\,\mathrm{s}^{-1}$) \\
            \hline
            $0.035$ & $13.87 \pm 3.64$ \\
            $0.056$ &  $6.18 \pm 4.86$ \\
            $0.090$ &  $7.57 \pm 4.21$ \\
            $0.109$ & $11.28 \pm 3.18$ \\
            $0.176$ &  $4.54 \pm 9.37$ \\
            $0.273$ &  $7.64 \pm 1.27$ \\
            \hline
        \end{tabular}
        \tablefoot{%
            The radii of the bins correspond to the average projected radius of the stars in each bin.%
        }
    \end{table}
    It is these binned dispersion measurements to which pyGravSphere fits, while CJAM fits directly to the unbinned velocity data in Table~\ref{tab:selection}.

    There are a number of models built into pyGravSphere to represent the density profiles of dark matter and stellar tracers and the velocity anisotropy profile.
    We choose to model the velocity anisotropy with the model of \citet{Baes-2007-A&A-471-419},
        \begin{equation}
            \beta_\mathrm{aniso}(r) = \beta_0 + (\beta_\infty - \beta_0) \frac{1}{1 + (r_0/r)^\eta},
            \label{eq:baes}
        \end{equation}
    which features a transition with rapidity~$\eta$ at radius~$r_0$ between an inner anisotropy~$\beta_0$ and an outer anisotropy~$\beta_\infty$.
    The anisotropy parameter is defined as
    \begin{equation}
        \beta_\mathrm{aniso}(r) := 1 - \frac{\sigma_\mathrm{t}^2(r)}{\sigma_\mathrm{r}^2(r)},
        \label{eq:aniso}
    \end{equation}
    where $\sigma_\mathrm{t}(r)$ and $\sigma_\mathrm{r}(r)$ are the tangential and radial component of the velocity dispersion, respectively.
    Here we will use the symmetrized anisotropy parameter~\citep{Read-2006-MNRAS-367-387},
    \begin{equation}
        \tilde{\beta}_\mathrm{aniso}(r) := \frac{\sigma_\mathrm{r}(r)-\sigma_\mathrm{t}(r)}{\sigma_\mathrm{r}(r)+\sigma_\mathrm{t}(r)} = \frac{\beta_\mathrm{aniso}(r)}{2-\beta_\mathrm{aniso}(r)},
        \label{eq:symaniso}
    \end{equation}
    which has the advantage of being bounded between $-1$ (fully tangential) and $+1$ (fully radial).
    Consequently, we define
    \begin{align}
       \tilde{\beta}_0 &\coloneqq \frac{\beta_0}{2-\beta_0},
       \label{eq:beta0}\\
       \tilde{\beta}_\infty &\coloneqq \frac{\beta_\infty}{2-\beta_\infty}.
       \label{eq:betainf}
    \end{align}
    We model the tracer profile with three \citet{Plummer-1911-MNRAS-71-460} profiles
    \begin{equation}
        \nu(r) = \sum_{j=1}^3 \frac{3M_j}{4\pi a_j^3} \Bigg(1 + \frac{r^2}{a_j^2}\Bigg)^{5/2}
        \label{eq:3plum}
    \end{equation}
    with masses~$M_j$ and radii~$a_j$.
    As pyGravSphere assumes spherical symmetry, a circular distribution is fit to the elliptical distribution on the sky.
    The dark-matter component can be modelled with a five-segment broken power-law profile~\citep{Read-2017-MNRAS-471-4541},
    \begin{equation}
        \rho_\text{pl}(r) =
        \begin{cases}
            \rho_0 (r/r_0)^{-\gamma_0}, & r < r_0, \\
            \rho_0 (r/r_{j+1})^{-\gamma_{j+1}} \prod_{n=0}^{n<j+1} (r_{n+1}/r_n)^{-\gamma_{n+1}}, & r_j < r < r_{j+1}, \\
        \end{cases}
        \label{eq:pl}
    \end{equation}
    or a Hernquist--Zhao \citep{Hernquist-1990-ApJ-356-359, Zhao-1996-MNRAS-278-488} profile,
    \begin{equation}
        \rho_\text{HZ}(r) = \frac{\rho_0}{(r/r_\mathrm{s})^\gamma (1 + (r/r_\mathrm{s})^\alpha)^{(\beta-\gamma)/\alpha}},
        \label{eq:zhao}
    \end{equation}
    also known as the $(\alpha, \beta, \gamma)$ profile.
    As a special case of the Hernquist--Zhao profile, we also look at the NFW profile with $(\alpha, \beta, \gamma) = (1, 3, 1)$, which is the same profile as for the CJAM CDM model.
    The broken power-law profile and Hernquist--Zhao profile allow for steeper slopes at large radii than the NFW/CDM, SIDM, and FDM models.
    Steep outer slopes can be a sign of stripping or truncation of the halo, for example due to tidal interactions with the Milky Way.
    The broken power-law profile should be especially suited for modelling truncated profiles because of its segmented nature.

    The pyGravSphere code uses emcee~\citep{ForemanMackey-2013-PASP-125-306} to constrain the parameter space.
    The use of this package, as well as the efficient implementations of the profile functions, makes pyGravSphere a fast code despite the high number of parameters it tries to constrain.
    Unfortunately, the use of an MCMC method makes comparison between models harder, as it does not readily provide Bayesian evidence.
    We remedy this by computing an approximation of the Bayesian evidence on the Markov chains with MCEvidence~\citep{2017arXiv170403472H}, using the estimator based on the nearest neighbours.

    Due to the limited quantity of data and the degeneracies between some of the parameters, we have extended some of the default pyGravSphere priors on the dark-matter parameters.
    We set the minimum value of $r_\text{s}$ to the projected radius of the innermost datum, rounded to the nearest decade, because we are not able to probe any smaller scales than the minimum radius.
    The maximum characteristic density is adjusted accordingly to not be a limiting bound.
    Conversely, we increase the maximum scale radius and decrease the minimum characteristic density.
    We increase the maximum allowed values of the Hernquist--Zhao $\beta$ parameter and power-law $\gamma_i$ to allow for steeper declines in density.
    For the same reason, we effectively remove the restriction on the difference between consecutive power-law slopes by setting the maximum difference between consecutive slopes equal to the difference between the prior minimum and maximum.
    Thus we effectively require only that the steepness of the broken power-law segments increases with the distance to the centre.
    An overview of the priors on the dark-matter parameters is given in Table~\ref{tab:pgspri}.
    \begin{table}
        \caption{Limits of the uniform pyGravSphere/emcee priors on the dark-matter parameters.}
        \label{tab:pgspri}
        \centering
        \begin{tabular}{lcc}
            \hline
            \hline
            Prior & Minimum & Maximum \\
            \hline
            $\log_{10}(\rho_0/M_\sun\,\mathrm{kpc}^{-3})$ & $3$ & $15$ \\
            $\log_{10}(r_\mathrm{s}/\mathrm{kpc})$ & $-2.5$ & $2.5$ \\
            $\alpha$\tablefootmark{(a)} & $0.5$ & $3$ \\
            $\beta$\tablefootmark{(a)} & $3$ & $9$ \\
            $\gamma$\tablefootmark{(a)} & $0$ & $1.5$ \\
            $\gamma_i$\tablefootmark{(b)} & $0$ & $9$ \\
            $\tilde\beta_0$ & $-1$ & $1$ \\
            $\tilde\beta_\infty$ & $-1$ & $1$ \\
            $\log_{10}(r_0/\mathrm{kpc})$ & $\log_{10}(0.5R_{1/2}/\mathrm{kpc})$ & $\log_{10}(2R_{1/2}/\mathrm{kpc})$ \\
            $\eta$ & $1$ & $3$ \\
            \hline
        \end{tabular}
        \tablefoot{%
            Listed are the characteristic density~$\rho_0$, the Navarro--Frenk--White~(NFW) scale radius~$r_\mathrm{s}$~(Eq.~\eqref{eq:cdm}), the Hernquist--Zhao $\alpha$, $\beta$, and $\gamma$ parameters~(Eq.~\eqref{eq:zhao}), the broken power-law slopes~$\gamma_i$~(Eq.~\eqref{eq:pl}), the symmetrized inner and outer velocity anisotropies~$\tilde\beta_0$ and~$\tilde\beta_\infty$~(Eq.~\eqref{eq:beta0}--\eqref{eq:betainf}), the anisotropy transition radius~$r_0$, and the sharpness~$\eta$ of the anisotropy transition~(Eq.~\eqref{eq:baes}).
            \tablefoottext{a}{In the case of the NFW model, $\alpha$, $\beta$, and $\gamma$ are fixed to $1$, $3$, and $1$, respectively.}
            \tablefoottext{b}{Within the indicated priors, $\gamma_{i+1} \geq \gamma_i$.}%
        }
    \end{table}

    We use the same settings for the MCMC walkers as \citet{2020MNRAS.498..144G}: $10^3$~walkers, making $2 \times 10^4$~steps, of which the first half is discarded as burn-in, and using $100$~integration points.
    Similarly, we analyse the resulting chains by first discarding samples with a $\chi^2$ more than ten times the minimum~$\chi^2$ and then drawing $10^5$ samples from the remaining samples.
    The best-fitting combination of parameters have a minimum $\chi^2$ less than $2$ for all three models, or a minimum reduced $\chi^2$ less than $1/3$, which indicates all models are good fits to the data.

\section{Results}
\label{sec:results}
    Using the two analysis methods presented above, we sample the parameter spaces of our dark matter--density profiles given the kinematical measurements of Eri~2.
    We break down the presentation of the results in several parts.
    In Sect.~\ref{ssec:param} we show the constraints on the density profiles and dark-matter models.
    This is followed by the presentation of the recovered density profiles in Sect.~\ref{ssec:recovery}, together with derived halo masses, concentrations, mass-to-light ratios, and astrophysical $J$~and $D$~factors.
    We then compare different dark-matter models using Bayesian evidence (Sect.~\ref{ssec:modcomp}).
    We remind the reader that each model is represented by the same colour in every figure.

\subsection{Parameter estimation}
\label{ssec:param}
\begin{figure}
        \includegraphics[width=\linewidth]{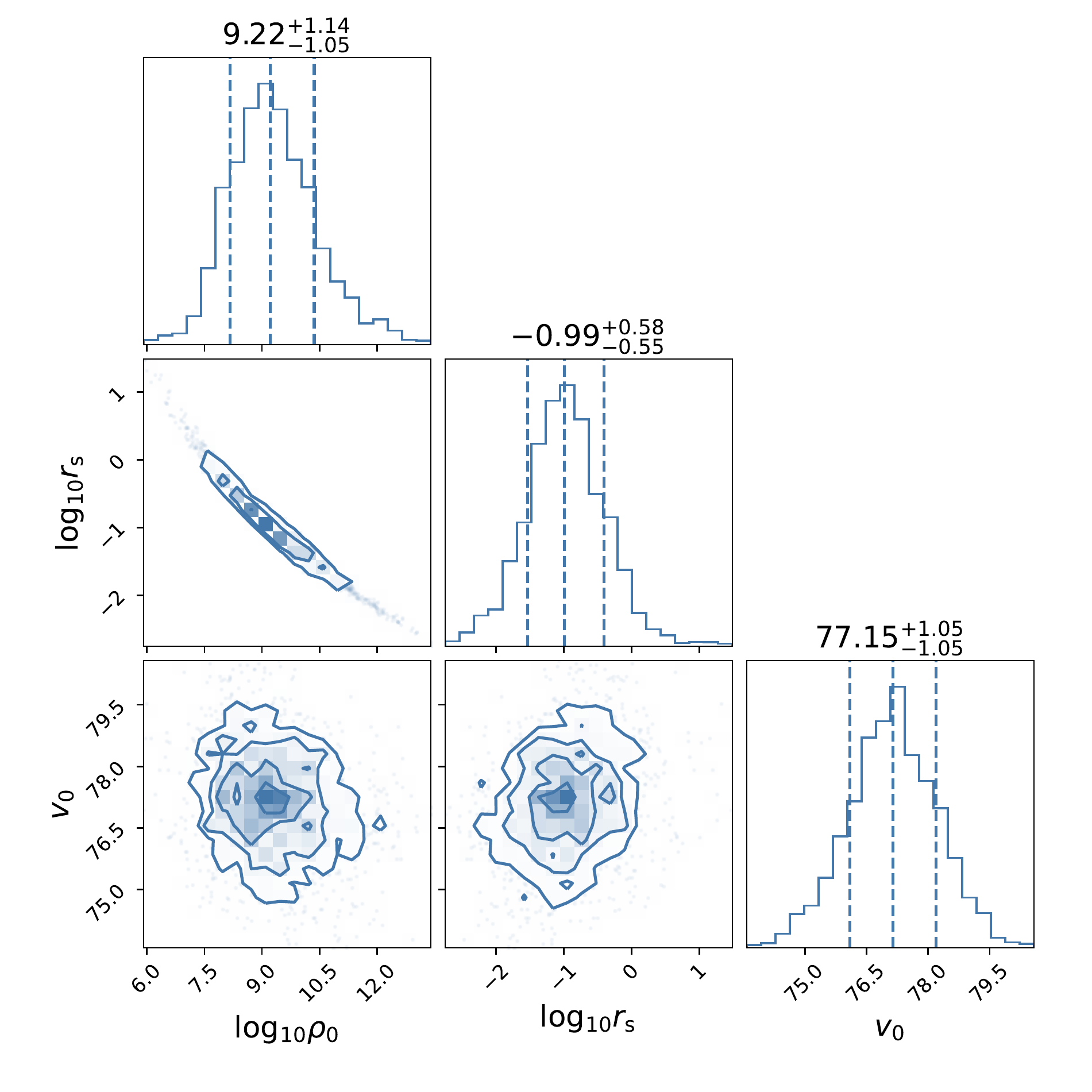}
        \caption{%
            Constraints on the dark matter--density profile of Eridanus~2 in the astrophysical parametrization, assuming cold dark matter, found using CJAM and MultiNest.
            Units are omitted for clarity.
            The parameters are the characteristic dark-matter density~$\rho_0$ in $M_\odot\,\mathrm{kpc}^{-3}$, the scale radius~$r_\mathrm{s}$ in $\mathrm{kpc}$, and the systemic velocity~$v_0$ in $\mathrm{km}\,\mathrm{s}^{-1}$.
            The contours correspond to $0.5\sigma$, $1.0\sigma$, $1.5\sigma$, and $2.0\sigma$ confidence levels, where $\sigma$ is the standard deviation of a two-dimensional normal distribution.
            The vertical dashed lines in the panels on the diagonal indicate the median and 68-\% confidence interval.%
        }
        \label{fig:cdmcorner}
    \end{figure}

\begin{figure}
        \includegraphics[width=\linewidth]{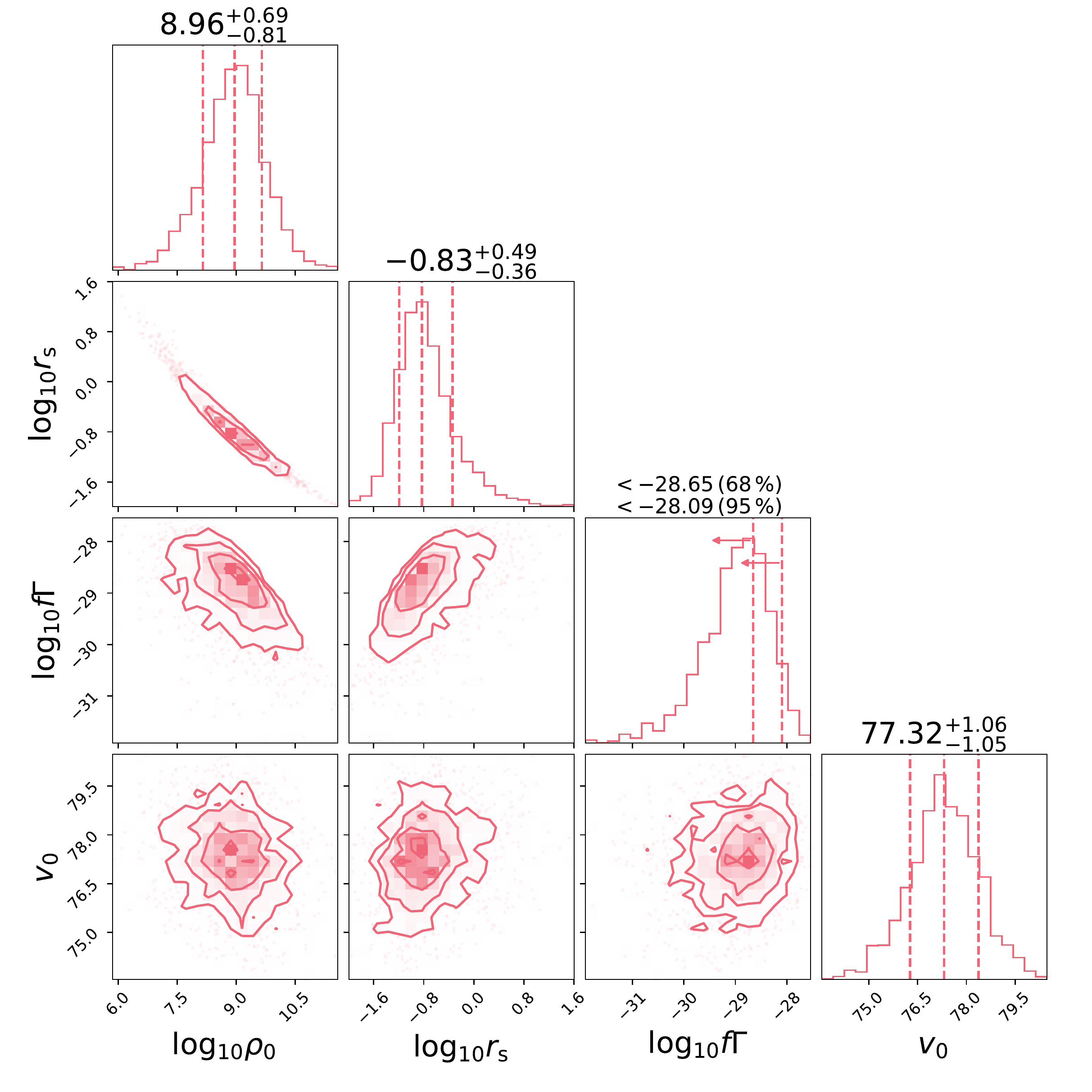}
        \caption{%
            Constraints on the dark matter--density profile of Eridanus~2 in the microphysical parametrization, assuming self-interacting dark matter, found using CJAM and MultiNest.
            Units are omitted for clarity.
            The parameters are the characteristic dark-matter density~$\rho_0$ in $M_\odot\,\mathrm{kpc}^{-3}$, the scale radius~$r_\mathrm{s}$ in $\mathrm{kpc}$, the effective self-interaction coefficient $f\Gamma$ in $\mathrm{cm}^3\,\mathrm{s}^{-1}\,\mathrm{eV}^{-1}\,c^2$, and the systemic velocity~$v_0$ in $\mathrm{km}\,\mathrm{s}^{-1}$.
            The contours correspond to $0.5\sigma$, $1.0\sigma$, $1.5\sigma$, and $2.0\sigma$ confidence levels, where $\sigma$ is the standard deviation of a two-dimensional normal distribution.
            The vertical dashed lines in the panels on the diagonal indicate the median and 68-\% confidence interval (without arrows) or the 68-\% and 95-\% confidence limits (upper and lower arrows, respectively).%
        }
        \label{fig:sidmcorner}
    \end{figure}

\begin{figure}
        \includegraphics[width=\linewidth]{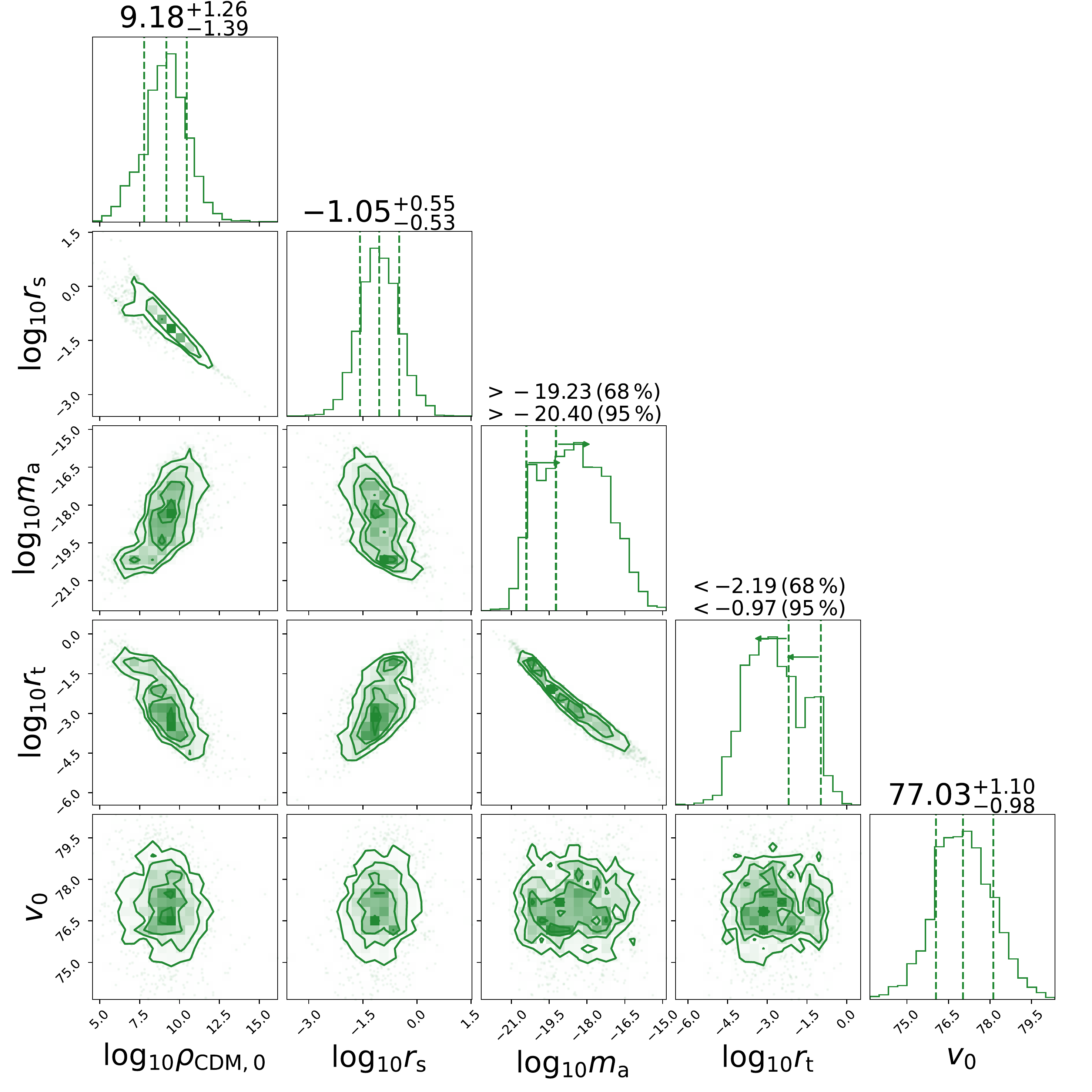}
        \caption{%
            Constraints on the dark matter--density profile of Eridanus~2 in the microphysical parametrization, assuming fuzzy dark matter, found using CJAM and MultiNest.
            Units are omitted for clarity.
            The parameters are the characteristic dark-matter density~$\rho_{\mathrm{CDM},0}$ of the CDM-like outer profile in $M_\odot\,\mathrm{kpc}^{-3}$, the scale radius~$r_\mathrm{s}$ of the CDM-like outer profile in $\mathrm{kpc}$, the dark matter--particle mass $m_\mathrm{a}$ in $\mathrm{eV}\,c^{-2}$, the transition radius $r_\mathrm{t}$ between the inner soliton and outer CDM-like profile in $\mathrm{kpc}$, and the systemic velocity~$v_0$ in $\mathrm{km}\,\mathrm{s}^{-1}$.
            The contours correspond to $0.5\sigma$, $1.0\sigma$, $1.5\sigma$, and $2.0\sigma$ confidence levels, where $\sigma$ is the standard deviation of a two-dimensional normal distribution.
            The vertical dashed lines in the panels on the diagonal indicate the median and 68-\% confidence interval (without arrows) or the 68-\% and 95-\% confidence limits (upper and lower arrows, respectively).%
        }
        \label{fig:fdm4corner}
    \end{figure}

    We show the constraints in the astrophysical parametrization of the CJAM CDM model in Figure~\ref{fig:cdmcorner} and the constraints in the microphysical parametrization of the SIDM and FDM models in Figures~\ref{fig:sidmcorner} and~\ref{fig:fdm4corner}, respectively.
    The constraints in the computational parametrizations for all three models and the astrophysical parametrizations for the SIDM and FDM models are displayed in Appendix~\ref{app:cjamsuppl}.
    Below we present and compare the constraints on the most important profile parameters.
    Quantities derived from the profiles, such as virial mass and concentration, will be presented in Section~\ref{ssec:recovery} together with the recovered profiles.

\paragraph{CJAM $\rho_0$ and $r_\mathrm{s}$}
    For the CDM profile we find a characteristic density of $\rho_0/(M_\sun\,\mathrm{kpc}^{-3}) = 10^{9.22^{+1.14}_{-1.05}} = 1.7^{+21.2}_{-1.5} \times 10^9$ and a scale radius of $r_\mathrm{s}/\mathrm{pc} = 10^{2.01^{+0.58}_{-0.55}} = 102^{+287}_{-73}$.
    The SIDM profile has consistent values for the same parameters: $\rho_0/(M_\sun\,\mathrm{kpc}^{-3}) = 10^{8.96^{+0.69}_{-0.81}} = 9.1^{+35.5}_{-7.7} \times 10^8$ and $r_\mathrm{s}/\mathrm{pc} = 10^{2.17^{+0.49}_{-0.36}} = 148^{+309}_{-83}$.
    This indicates that at large radii the density profiles of CDM and SIDM are in agreement.

\paragraph{CJAM SIDM $r_\mathrm{c}$ and $f\Gamma$}
    Considering the SIDM core radius is consistent with a scale radius smaller than our smallest projected radius ($1.96\,\mathrm{pc}$), we lack constraining power at the lower end of the range of this parameter.
    It is therefore appropriate to present the constraint as an upper limit: $r_\mathrm{c}/\mathrm{pc} < 10^{1.67} = 47$ at the 68-\% confidence level and $r_\mathrm{c}/\mathrm{pc} < 10^{2.07} = 117$ at the 95-\% confidence level.
    For the related effective self-interaction coefficient we find that $f\Gamma/(\mathrm{cm}^3\,\mathrm{s}^{-1}\,\mathrm{eV}^{-1}\,c^2) < 10^{-28.65} = 2.2 \times 10^{-29}$ at the 68-\% confidence level and $f\Gamma/(\mathrm{cm}^3\,\mathrm{s}^{-1}\,\mathrm{eV}^{-1}\,c^2) < 10^{-28.09} = 8.1 \times 10^{-29}$ at the 95-\% confidence level.

\paragraph{CJAM FDM $r_\mathrm{sol}$ and $m_\mathrm{a}$}
    In the case of the FDM model, it is also appropriate to present the soliton radius as an upper limit: $r_\mathrm{sol}/\mathrm{pc} < 10^{0.86} = 7.2$ at the 68-\% confidence level and $r_\mathrm{sol}/\mathrm{pc} < 10^{2.01} = 102$ at the 95-\% confidence level.
    Because of the degeneracy between the soliton radius and central soliton density, the central soliton density should then be understood as a lower limit: $\rho_{\mathrm{sol},0}/(M_\sun\,\mathrm{kpc}^{-3}) > 10^{11.89} = 7.8 \times 10^{11}$ at the 68-\% confidence level and $\rho_{\mathrm{sol},0}/(M_\sun\,\mathrm{kpc}^{-3}) > 10^{10.13} = 1.3 \times 10^{10}$ at the 95-\% confidence level.
    The equivalent dark matter--particle mass is given as $m_\mathrm{a}/(\mathrm{eV}\,c^{-2}) > 10^{-19.23} = 5.9 \times 10^{-20}$ at the 68-\% confidence level and $m_\mathrm{a}/(\mathrm{eV}\,c^{-2}) > 10^{-20.40} = 4.0 \times 10^{-21}$ at the 95-\% confidence level.

\begin{figure*}
        \includegraphics[width=\linewidth]{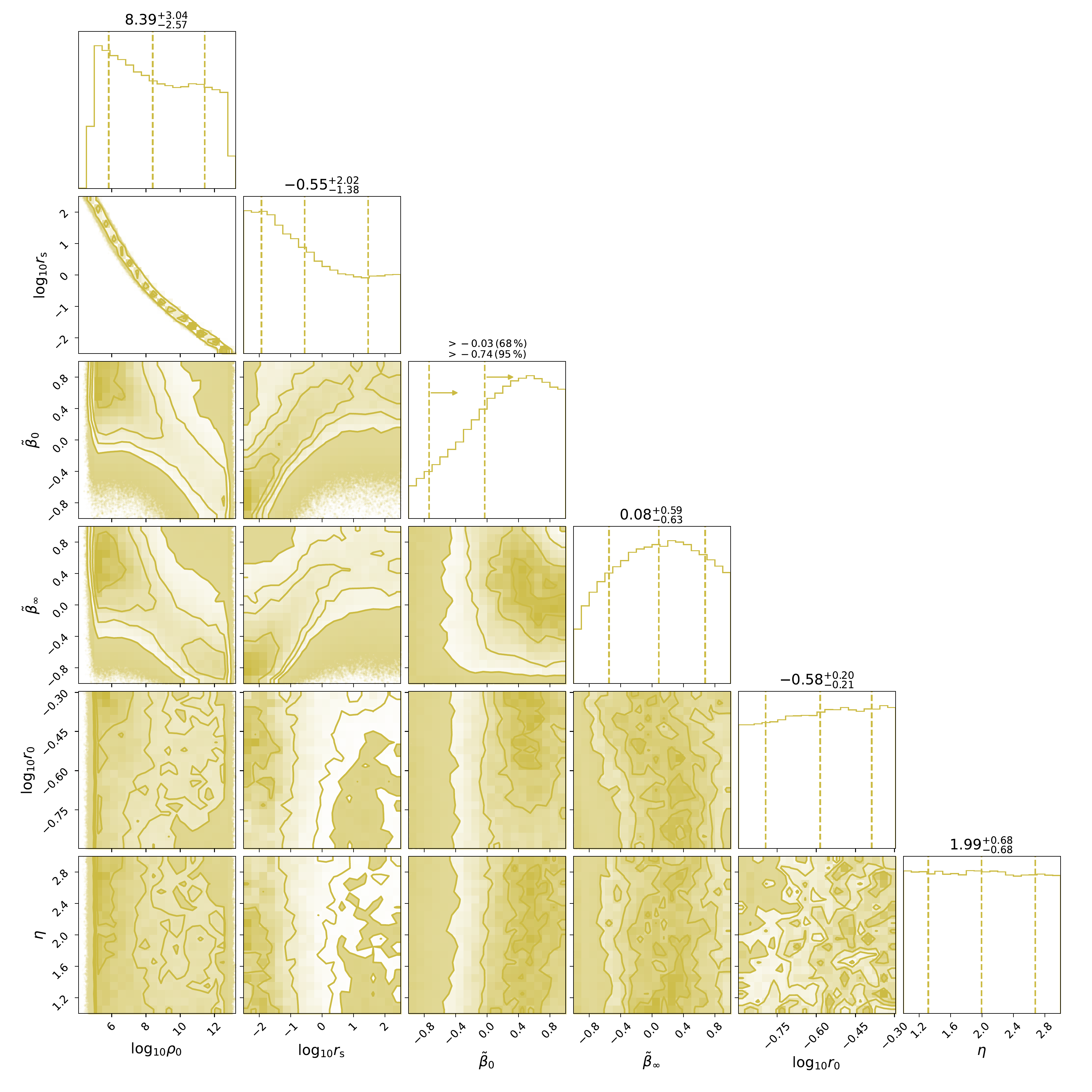}
        \caption{%
            Constraints on the dark matter--density profile of Eridanus~2, assuming a Navarro--Frenk--White profile, found using pyGravSphere.
            Units are omitted for clarity.
            The parameters are the characteristic dark-matter density~$\rho_0$ in $M_\odot\,\mathrm{kpc}^{-3}$, the scale radius~$r_\mathrm{s}$ in $\mathrm{kpc}$, the symmetrized inner and outer velocity anisotropy~$\tilde\beta_0$ and~$\tilde\beta_\infty$, the transition radius~$r_0$ between inner and outer velocity anisotropy in $\mathrm{kpc}$, and the sharpness~$\eta$ of the velocity-anisotropy transition.
            The contours correspond to $0.5\sigma$, $1.0\sigma$, $1.5\sigma$, and $2.0\sigma$ confidence levels, where $\sigma$ is the standard deviation of a two-dimensional normal distribution.
            The vertical dashed lines in the panels on the diagonal indicate the median and 68-\% confidence interval (without arrows) or the 68-\% and 95-\% confidence limits (upper and lower arrows, respectively).%
        }
        \label{fig:nfwcorner}
    \end{figure*}

\begin{figure*}
        \includegraphics[width=\linewidth]{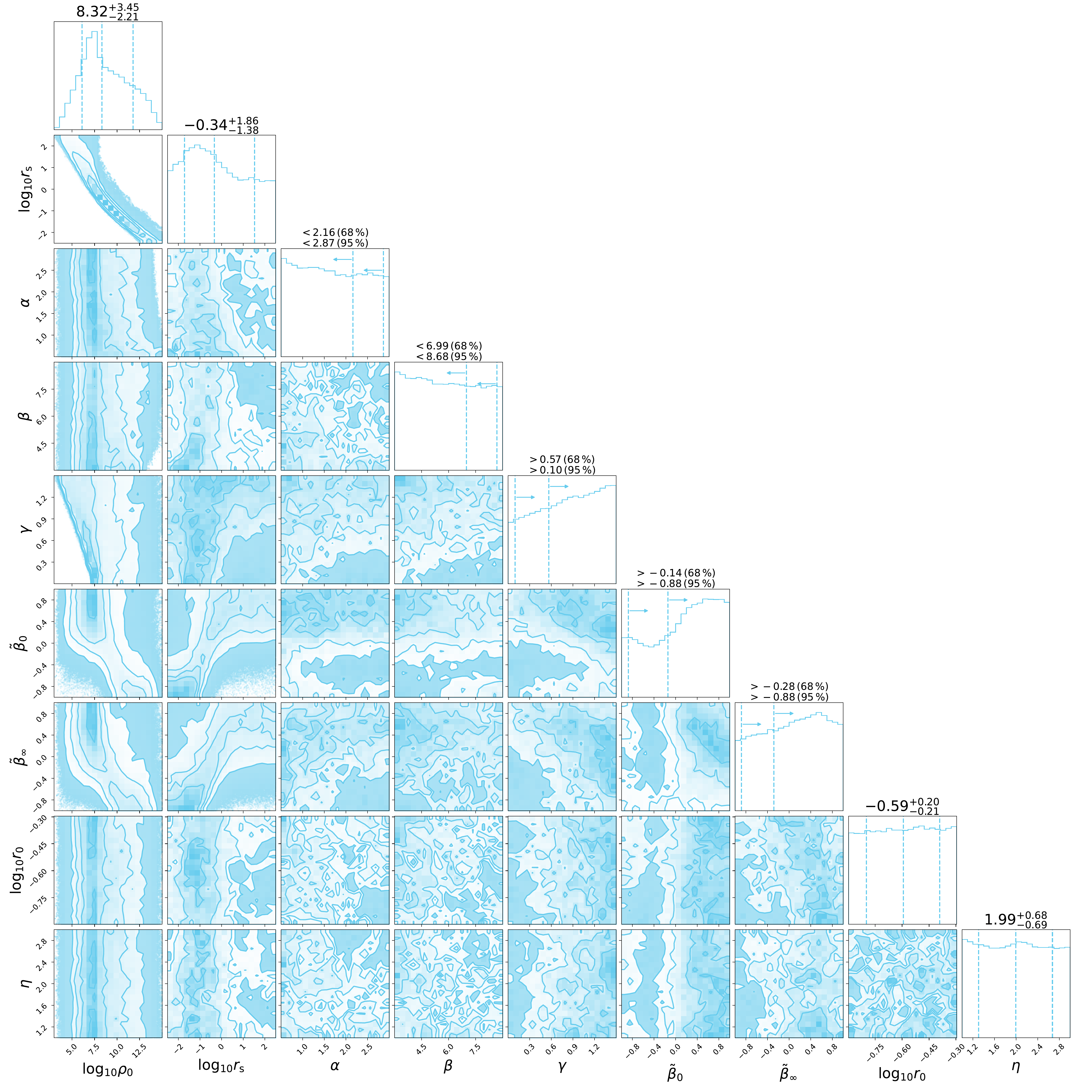}
        \caption{%
            Constraints on the dark matter--density profile of Eridanus~2, assuming a Hernquist--Zhao profile, found using pyGravSphere.
            Units are omitted for clarity.
            The parameters are the characteristic dark-matter density~$\rho_0$ in $M_\odot\,\mathrm{kpc}^{-3}$, the scale radius~$r_\mathrm{s}$ in $\mathrm{kpc}$, the inner and outer negative logarithmic slopes~$\gamma$ and~$\beta$ and the sharpness~$\alpha$ of their transition, the symmetrized inner and outer velocity anisotropy~$\tilde\beta_0$ and~$\tilde\beta_\infty$, the transition radius~$r_0$ between inner and outer velocity anisotropy in $\mathrm{kpc}$, and the sharpness~$\eta$ of the velocity-anisotropy transition.
            The contours correspond to $0.5\sigma$, $1.0\sigma$, $1.5\sigma$, and $2.0\sigma$ confidence levels, where $\sigma$ is the standard deviation of a two-dimensional normal distribution.
            The vertical dashed lines in the panels on the diagonal indicate the median and 68-\% confidence interval (without arrows) or the 68-\% and 95-\% confidence limits (upper and lower arrows, respectively).%
        }
        \label{fig:zhaocorner}
    \end{figure*}

\begin{figure*}
        \includegraphics[width=\linewidth]{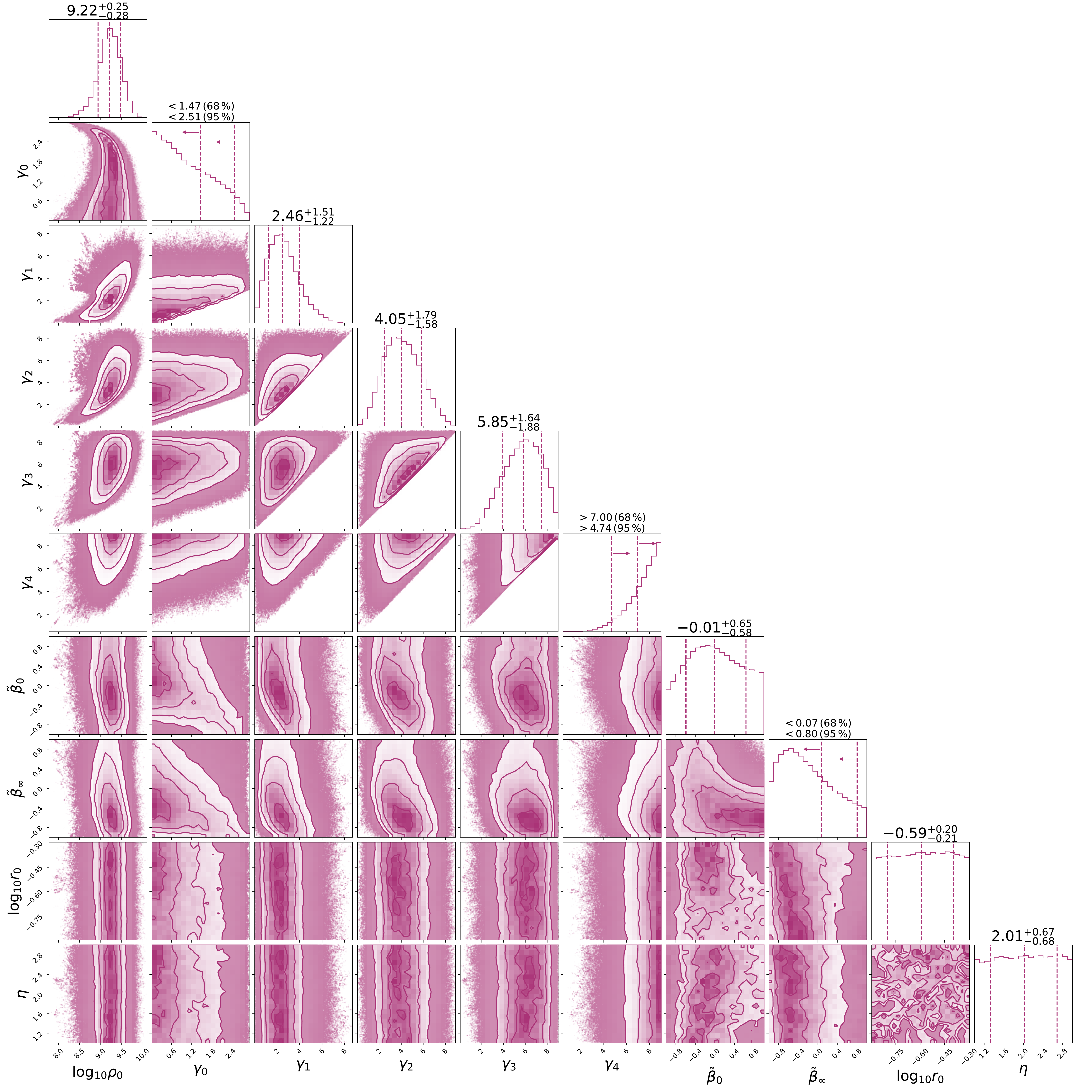}
        \caption{%
            Constraints on the dark matter--density profile of Eridanus~2, assuming a broken power-law profile, found using pyGravSphere.
            Units are omitted for clarity.
            The parameters are the characteristic dark-matter density~$\rho_0$ in $M_\odot\,\mathrm{kpc}^{-3}$, the negative power-law slopes $\gamma_0, \ldots, \gamma_4$, the symmetrized inner and outer velocity anisotropy~$\tilde\beta_0$ and~$\tilde\beta_\infty$, the transition radius~$r_0$ between inner and outer velocity anisotropy in $\mathrm{kpc}$, and the sharpness~$\eta$ of the velocity-anisotropy transition.
            The contours correspond to $0.5\sigma$, $1.0\sigma$, $1.5\sigma$, and $2.0\sigma$ confidence levels, where $\sigma$ is the standard deviation of a two-dimensional normal distribution.
            The vertical dashed lines in the panels on the diagonal indicate the median and 68-\% confidence interval (without arrows) or the 68-\% and 95-\% confidence limits (upper and lower arrows, respectively).%
        }
        \label{fig:plnscorner}
    \end{figure*}

\paragraph{pyGravSphere}
    Figures~\ref{fig:nfwcorner}, \ref{fig:zhaocorner}, and \ref{fig:plnscorner} show the parameter constraints for the pyGravSphere NFW, Hernquist--Zhao, and broken power-law models, respectively.
    The characteristic density of the NFW model is $\rho_0/(M_\sun\,\mathrm{kpc}^{-3}) = 10^{8.39^{+3.04}_{-2.57}} = 0.25^{+268.91}_{-0.24} \times 10^9$ and its scale radius is $r_\mathrm{s}/\mathrm{pc} = 10^{2.45^{+2.02}_{-1.38}} = 282^{+29230}_{-270}$, consistent with the CJAM CDM results, but also strongly degenerate.
    For the Hernquist--Zhao model we find that $\rho_0/(M_\sun\,\mathrm{kpc}^{-3}) = 10^{8.32^{+3.45}_{-2.21}} = 0.21^{+588.63}_{-0.21} \times 10^9$ and $r_\mathrm{s}/\mathrm{pc} = 10^{2.66^{+1.86}_{-1.38}} = 457^{+32656}_{-438}$, which is again consistent but degenerate.
    The characteristic density of the broken power-law model, $\rho_0/(M_\sun\,\mathrm{kpc}^{-3}) = 10^{9.22^{+0.25}_{-0.28}} = 1.66^{+1.29}_{-0.79}$, is not directly comparable to the other characteristic densities due to the difference in the definitions, but it is notable that this parameter is much better constrained.
    The Hernquist--Zhao model prefers inner slopes $\gamma > 0.57$ at the 68-\% confidence level and $\gamma > 0.10$ at the 95-\% confidence level that are consistent with a cusp, while the broken power-law model has a weak preference for a core with $\gamma_0 < 1.47$ at the 68-\% confidence level and $\gamma_0 < 2.51$ at the 95-\% confidence level, but is still also consistent with a cusp.
    Conversely, the Hernquist--Zhao model weakly prefers outer slopes consistent with CDM, with $\beta < 6.99$ at the 68-\% confidence level and $\beta < 8.68$ at the 95-\% confidence level, while the broken power-law model prefers steeper slopes with $\gamma_4 > 7.00$ at the 68-\% confidence level and $\gamma_4 > 4.74$ at the 95-\% confidence level.
    The shape of the Hernquist--Zhao profile is thus consistent with the NFW profile, albeit with large uncertainty, while the shape of the broken power-law profile deviates at large radii by over $2\sigma$.
    The constraints on the velocity anisotropies are in general very weak, with an apparent trend for positive (radial) anisotropy in the case of the Hernquist--Zhao profile and for the centre in the case of the NFW profile.
    At large radii the NFW profile seems to prefer isotropy.
    The broken power-law model profile, on the other hand, prefers isotropy for the centre and negative (tangential) anisotropy for the outer radii.
    The transition between these possibly different regimes of inner and outer velocity anisotropy is essentially unconstrained.

\subsection{Profile recovery}
\label{ssec:recovery}
\begin{figure}
        \includegraphics[width=\linewidth]{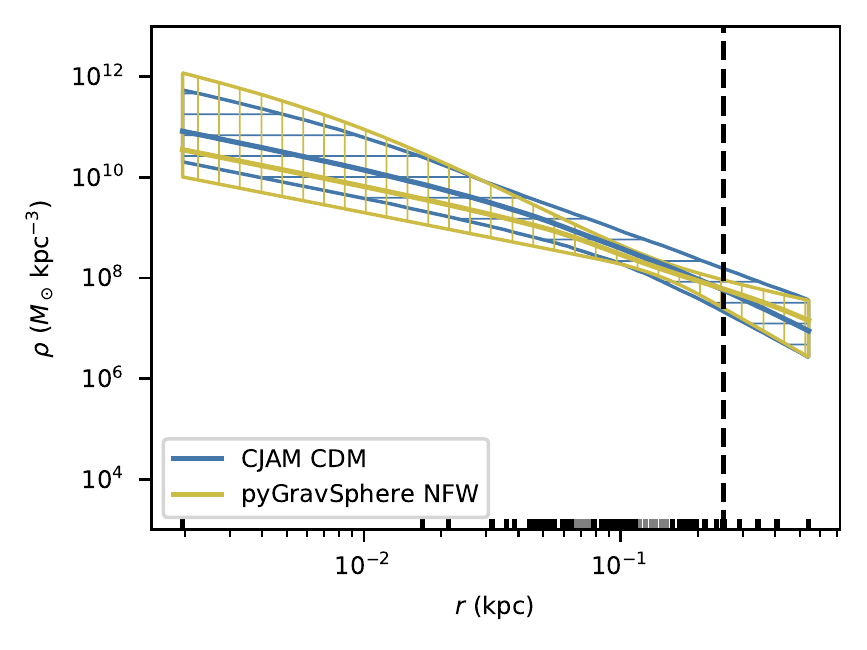}
        \caption{%
            Recovered dark matter--density profile of Eridanus~2, comparing the CJAM model for cold dark matter with the PyGravSphere Navarro--Frenk--White profile.
            These models have the same functional form for the density profile but use different assumptions and methods of calculation.
            The hatched bands represent the 68-\% confidence interval on the density at each radius.
            The half-light radius is indicated with the vertical dashed line.
            The black markers at the bottom of the figure show the projected radii of the kinematic tracers.
            Tracers in bins rejected by pyGravSphere are marked in grey.%
        }
        \label{fig:cdmnfw}
    \end{figure}

    The two methods, CJAM and pyGravSphere, that we use to constrain the density profile of Eri~2 have one profile model in common: the CDM/NFW profile.
    By comparing the constraints on this profile model obtained with the two methods, we can gauge the influence of the different assumptions that go into the methods.
    In Figure~\ref{fig:cdmnfw} we show the recovered CDM/NFW density profiles as a function of radius in the form of the median density and 68-\% confidence interval at every radius.
    Although there are differences, most noticeably that pyGravSphere prefers lower central densities and higher outer densities than CJAM, the overall agreement is good.
    The two recovered profiles agree within the uncertainties at every radius and there is no systematic preference for higher or lower densities.
    This indicates that the different assumptions have no significant effect on the recovered constraints and lends support to the results of both methods.

\begin{figure*}
        \includegraphics[width=0.5\linewidth]{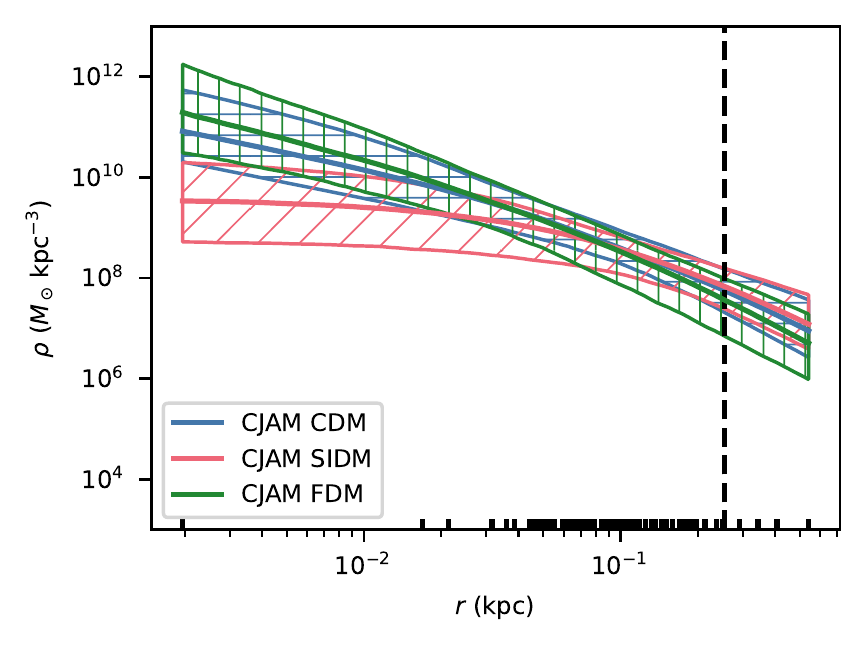}%
        \includegraphics[width=0.5\linewidth]{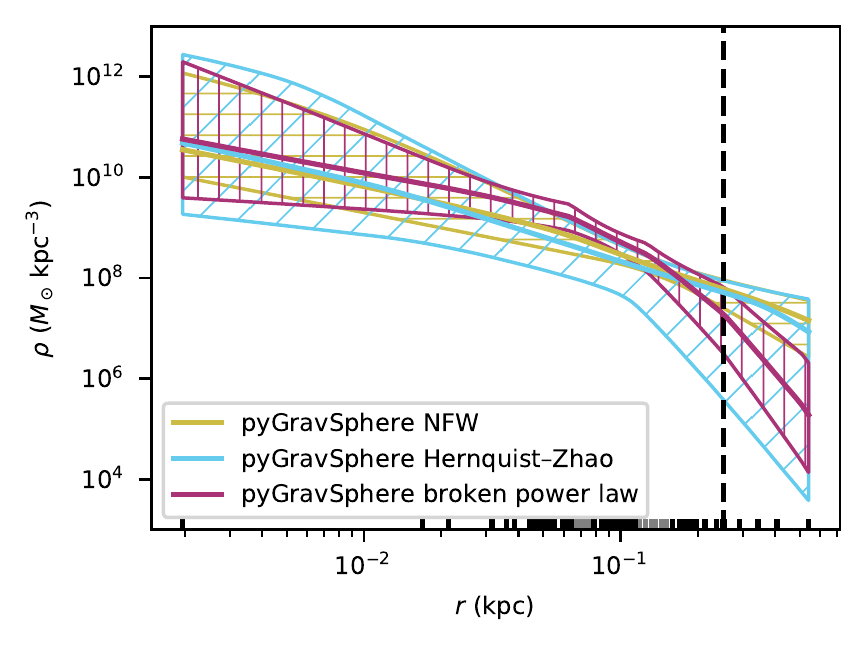}
        \caption{%
            Recovered dark matter--density profiles of Eridanus~2.
            (\textit{left}) CJAM models for cold dark matter~(CDM), self-interacting dark matter~(SIDM), and fuzzy dark matter~(FDM).
            (\textit{right}) pyGravSphere models with Navarro--Frenk--White~(NFW), Hernquist--Zhao, and broken power-law profiles.
            The hatched bands represent the 68-\% confidence interval on the density at each radius.
            The half-light radius is indicated with the vertical dashed line.
            The black markers at the bottom of the figure show the projected radii of the kinematic tracers.
            Tracers in bins rejected by pyGravSphere are marked in grey.%
        }
        \label{fig:recovery}
    \end{figure*}

    The recovered profiles using all models are displayed in Figure~\ref{fig:recovery}.
    Around the radius where we have the largest number of tracers, the agreement between the profiles is the best and the uncertainties are the smallest.
    At larger radii, five of the models agree very well, but the broken power-law model prefers lower densities in its last bin.
    This lower density could be an indication of the effect of tidal truncation, but the data are insufficient to conclude this, as we will show below.
    The disagreement is the largest at small radii, where the density at the projected position of the innermost tracer varies from ${\sim}10^{9.5}\,M_\sun\,\mathrm{kpc}^{-3}$ to ${\sim}10^{11.5}\,M_\sun\,\mathrm{kpc}^{-3}$.
    This is not surprising, considering the lack of tracers at these radii and that some models by design have more freedom at small radii.
    All profiles are in agreement at the smaller radii considering their uncertainties.
    In Appendix~\ref{app:vdisp} we show the recovered intrinsic velocity dispersion profiles and compare them to estimates directly derived from the measured line-of-sight velocities.

\begin{figure*}
        \includegraphics[width=0.5\linewidth]{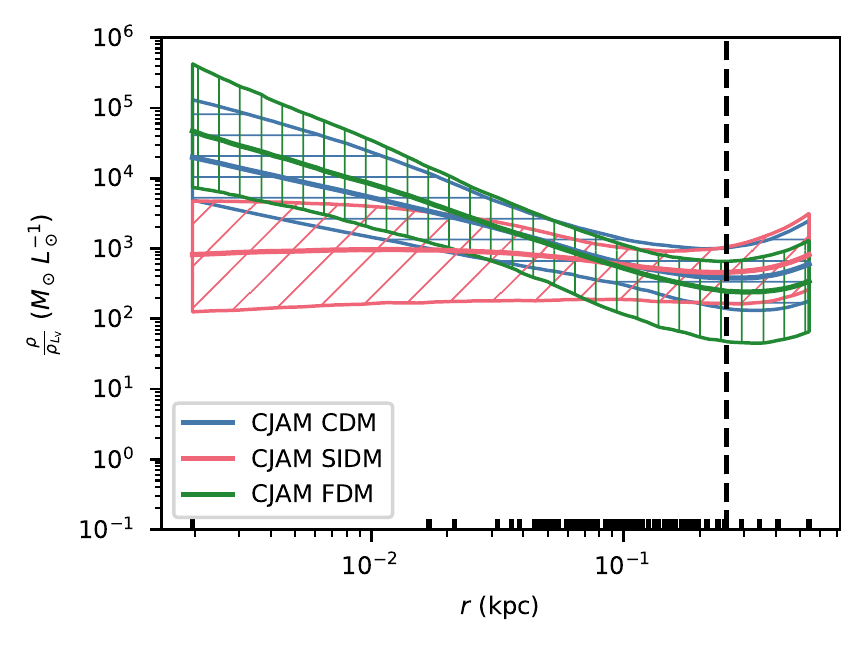}%
        \includegraphics[width=0.5\linewidth]{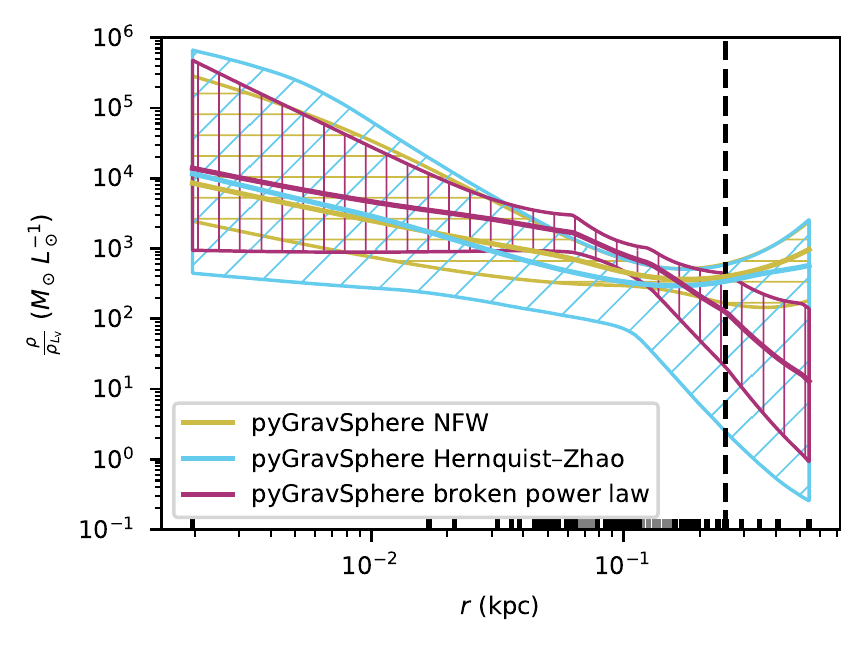}
        \caption{%
            Recovered de-projected mass-to-light profiles of Eridanus~2.
            These profiles show the local ratio of dark-matter density over luminosity density as a function of radius.
            The light profile is the exponential profile determined by \citet{Crnojevic-2016-ApJL-824-L14}.
            (\textit{left}) CJAM models for cold dark matter~(CDM), self-interacting dark matter~(SIDM), and fuzzy dark matter~(FDM).
            (\textit{right}) pyGravSphere models with Navarro--Frenk--White~(NFW), Hernquist--Zhao, and broken power-law dark-matter profiles.
            The hatched bands represent the 68-\% confidence interval on the mass-to-light ratio at each radius.
            The half-light radius is indicated with the vertical dashed line.
            The black markers at the bottom of the figure show the projected radii of the kinematic tracers.
            Tracers in bins rejected by pyGravSphere are marked in grey.%
        }
        \label{fig:mtol}
    \end{figure*}

    We display the local mass-to-light ratio as a function of radius in Figure~\ref{fig:mtol}.
    The density profile is divided by the V-band luminosity density profile, computed by de-projecting the exponential surface brightness profile from \citet{Crnojevic-2016-ApJL-824-L14} using the equation derived by \citet{Baes-2011-A&A-525-A136}.
    This is a local, three-dimensional mass-to-light ratio at the indicated radius, not a cumulative mass-to-light ratio within that radius.
    Since the luminosity density profile is the same for every dark-matter model, the same differences are visible between the models.
    For most models the local mass-to-light radius has a minimum of ${\sim}10^3\,M_\sun\,L_\sun^{-1}$ around the half-light radius.

    We compute virial and half-light quantities and the maximum circular velocity from the density profiles and list them in Tables~\ref{tab:m200cjam} and~\ref{tab:m200pgs} for CJAM and pyGravSphere profiles, respectively.
    \begin{table}
        \caption{Quantities derived from the CJAM/MultiNest density or mass profiles of Eridanus~2 under the assumption of different profile models.}
        \label{tab:m200cjam}
        \centering
        \begin{tabular}{lccc}
            \hline
            \hline
            Quantity & CDM & SIDM & FDM \\
            \hline
            $\log_{10}(r_{200}/\mathrm{kpc})$ & $0.96^{+0.20}_{-0.12}$ & $1.02^{+0.22}_{-0.11}$ & $0.88^{+0.17}_{-0.12}$ \\
            $\log_{10}(c_{200})$ & $1.95^{+0.40}_{-0.39}$ & $1.84^{+0.29}_{-0.29}$ & --- \\
            $\log_{10}(M_{200}/M_\sun)$ & $7.89^{+0.60}_{-0.36}$ & $8.07^{+0.64}_{-0.33}$ & $7.67^{+0.50}_{-0.36}$ \\
            $\log_{10}(\Upsilon_{200}/(M_\sun\,L_\sun^{-1}))$ & $3.12^{+0.60}_{-0.36}$ & $3.31^{+0.64}_{-0.33}$ & $2.90^{+0.50}_{-0.36}$ \\
            $\log_{10}(V_\mathrm{max}/(\mathrm{km}\,\mathrm{s}^{-1}))$ & $1.19^{+0.09}_{-0.06}$ & $1.19^{+0.10}_{-0.07}$ & $1.20^{+0.13}_{0.07}$ \\
            $\log_{10}(M_{1/2}/M_\sun)$ & $7.05^{+0.10}_{-0.11}$ & $7.07^{+0.10}_{-0.10}$ & $6.99^{+0.50}_{0.36}$ \\
            $\log_{10}(\Upsilon_{1/2}/(M_\sun\,L_\sun^{-1}))$ & $2.59^{+0.10}_{-0.11}$ & $2.61^{+0.10}_{-0.10}$ & $2.53^{+0.12}_{-0.17}$ \\
            $\log_{10}(J(\alpha_\mathrm{c}^J)/(M_\sun^2\,\mathrm{kpc}^{-5}))$ & $10.94^{+0.57}_{-0.38}$ & $10.66^{+0.31}_{-0.22}$ & $11.20^{+0.69}_{-0.51}$ \\
            $\log_{10}(D(\alpha_\mathrm{c}^D)/(M_\sun\,\mathrm{kpc}^{-2}))$ & $2.31^{+0.34}_{-0.22}$ & $2.42^{+0.35}_{-0.20}$ & $2.17^{+0.28}_{-0.25}$ \\
            \hline
        \end{tabular}
        \tablefoot{%
            The models are the cold dark matter~(CDM), self-interacting dark matter~(SIDM), and fuzzy dark matter~(FDM) profiles.
            Listed are the virial radius~$r_{200}$, the concentration parameter~$c_{200}$ (not defined for the FDM profile), the virial mass~$M_{200}$, the virial mass-to-light ratio~$\Upsilon_{200}$, the maximum circular velocity~$V_\mathrm{max}$, the half-light mass~$M_{1/2}$, the half-light mass-to-light ratio~$\Upsilon_{1/2}$, and the astrophysical $J$~and $D$~factors using the critical integration angles.%
        }
    \end{table}
    \begin{table}
        \caption{Quantities derived from the pyGravSphere/emcee density or mass profiles of Eridanus~2 under the assumption of different profile models.}
        \label{tab:m200pgs}
        \centering
        \begin{tabular}{lccc}
            \hline
            \hline
            Quantity & NFW & HZ & BPL \\
            \hline
            $\log_{10}(r_{200}/\mathrm{kpc})$ & $1.13^{+0.97}_{-0.30}$ & $0.89^{+1.13}_{-0.22}$ & $0.72^{+0.07}_{-0.05}$ \\
            $\log_{10}(c_{200})$ & $1.85^{+1.19}_{-1.00}$ & $1.51^{+1.27}_{-0.95}$ & --- \\
            $\log_{10}(M_{200}/M_\sun)$ & $8.41^{+2.90}_{-0.91}$ & $7.69^{+3.41}_{-0.66}$ & $7.17^{+0.19}_{-0.15}$ \\
            $\log_{10}(\Upsilon_{200}/(M_\sun\,L_\sun^{-1}))$ & $3.65^{+2.90}_{-0.91}$ & $2.93^{+3.41}_{-0.66}$ & $2.41^{+0.19}_{-0.15}$ \\
            $\log_{10}(V_\mathrm{max}/(\mathrm{km}\,\mathrm{s}^{-1}))$ & $1.34^{+0.61}_{-0.16}$ & $1.47^{+0.53}_{-0.26}$ & $1.26^{+0.11}_{-0.12}$ \\
            $\log_{10}(M_{1/2}/M_\sun)$ & $7.01^{+0.14}_{-0.16}$ & $6.99^{+0.14}_{-0.20}$ & $7.10^{+0.11}_{-0.12}$ \\
            $\log_{10}(\Upsilon_{1/2}/(M_\sun\,L_\sun^{-1}))$ & $2.55^{+0.14}_{-0.16}$ & $2.53^{+0.14}_{-0.20}$ & $2.64^{+0.11}_{-0.12}$ \\
            $\log_{10}(J(\alpha_\mathrm{c}^J)/(M_\sun^2\,\mathrm{kpc}^{-5}))$ & $10.91^{+1.31}_{-0.44}$ & $11.24^{+2.02}_{-0.71}$ & $11.25^{+1.20}_{-0.59}$ \\
            $\log_{10}(D(\alpha_\mathrm{c}^D)/(M_\sun\,\mathrm{kpc}^{-2}))$ & $2.57^{+1.56}_{-0.46}$ & $2.34^{+1.74}_{-0.50}$ & $2.03^{+0.16}_{-0.17}$ \\
            \hline
        \end{tabular}
        \tablefoot{%
            The models are the Navarro--Frenk--White~(NFW), Hernquist--Zhao~(HZ), and broken power-law~(BPL) profiles.
            Listed are the virial radius~$r_{200}$, the concentration parameter~$c_{200}$ (not defined for the broken power-law profile), the virial mass~$M_{200}$, the virial mass-to-light ratio~$\Upsilon_{200}$, the maximum circular velocity~$V_\mathrm{max}$, the half-light mass~$M_{1/2}$, the half-light mass-to-light ratio~$\Upsilon_{1/2}$, and the astrophysical $J$~and $D$~factors using the critical integration angles.%
        }
    \end{table}
    There is good agreement between the different profiles and between CJAM and pyGravSphere for the maximum circular velocity~$V_\mathrm{max}$ and for the mass $M_{1/2}$ within the projected half-light radius, and as a consequence also for the integrated mass-to-light ratio $\Upsilon_{1/2} = M_{1/2}/(L_\mathrm{V}/2)$ within the same radius.
    The virial mass~$M_{200}$ and mass-to-light ratio $\Upsilon_{200} = M_{200}/L_\mathrm{V}$ are more divergent from model to model.
    This is a consequence of the virial radius~$r_{200}$ being an order of magnitude larger than the projected radius of the outermost tracer.
    For the calculation of the virial quantities, the density profiles are extrapolated to an extent that a small change in the profile slope around the outermost tracer leads to a large difference in the virial radius and virial mass.
    From the virial mass, the V-band luminosity, and the stellar mass-to-light ratio of $1.56$ derived in \citetalias{Zoutendijk-2020-A&A-635-A107}, we can estimate a stellar-mass--to--halo-mass ratio of ${\sim}10^{-3}$.
    For this value a galaxy is expected to reside in a halo that is intermediate between cuspy and cored~\citep{DiCintio-2014-MNRAS-441-2986}.

    We also list in Tables~\ref{tab:m200cjam} and~\ref{tab:m200pgs} the astrophysical factors~$J$ and~$D$, which are used to calculate the (gamma-ray) flux from annihilation and decay, respectively, of dark-matter particles~\citep{Bergstroem-1998-APh-9-137}.
    These are integrals of the density profile or its square, over the line-of-sight~$l$ and a solid angle~$\Delta\Omega$ in the plane of the sky:
    \begin{align}
        J(\alpha) &= \int_{\Delta\Omega(<\alpha)} \int_{-\infty}^{+\infty} \rho^2\, \mathrm{d}l\, \mathrm{d}\Omega,
        \label{eq:jfactor}\\
        D(\alpha) &= \int_{\Delta\Omega(<\alpha)} \int_{-\infty}^{+\infty} \rho\, \mathrm{d}l\, \mathrm{d}\Omega.
        \label{eq:dfactor}
    \end{align}
    We calculate these integrals up to the critical integration angle, which is the planar angle corresponding to the circular solid angle for which these factors are found to be most constrained for dwarf spheroidal galaxies.
    The critical integration angle is the angle subtended by the half-light radius for the $D$~factor ($\alpha_\mathrm{c}^D = \mathrm{R}_{1/2}/D$)~\citep{Bonnivard-2015-MNRAS-446-3002} and twice the half-light radius for the $J$~factor ($\alpha_\mathrm{c}^J = 2\mathrm{R}_{1/2}/D$)~\citep{Walker-2011-ApJL-733-L46}.
    The $J$ and $D$ factors are generally consistent within their uncertainties, though there is some tension for the $D$ factor between the SIDM and broken-power law models.

\subsection{Model comparison}
\label{ssec:modcomp}
    We have so far placed constraints on astrophysical and microphysical parameters assuming different models and informally compared the different models based on the recovered profiles.
    The next question to ask, is which model provides the best fit to the data, which may indicate a preference for one form of dark matter over another.
    In Tables~\ref{tab:cjamev} and \ref{tab:pgsev} we present the Bayesian evidence~$Z$ for the CJAM and pyGravSphere models, respectively.
    \begin{table}
        \caption{Bayesian evidence comparison for CJAM/MultiNest models.}
        \label{tab:cjamev}
        \centering
        \begin{tabular}{lcc}
            \hline
            \hline
            Model & $\ln(Z)$ & $\Delta\log_{10}(Z)$ \\
            \hline
            CDM & $-360.9$ & $-0.4$ \\
            SIDM & $-362.3$ & $-1.0$ \\
            FDM & $-360.0$ & $0$ \\
            \hline
        \end{tabular}
        \tablefoot{%
            The models are cold dark matter~(CDM), self-interacting dark matter~(SIDM), and fuzzy dark matter~(FDM).
            For each model the natural logarithm of the Bayesian evidence and the decimal logarithm of the Bayes factor are shown.%
        }
    \end{table}
    \begin{table}
        \caption{Bayesian evidence comparison for pyGravSphere/emcee models using MCEvidence.}
        \label{tab:pgsev}
        \centering
        \begin{tabular}{lcc}
            \hline
            \hline
            Model & $\ln(Z)$ & $\Delta\log_{10}(Z)$ \\
            \hline
            NFW & $-101.2$ & $-0.7$ \\
            HZ & $-101.2$ & $-0.7$ \\
            BPL & $-99.6$ & $0$ \\
            \hline
        \end{tabular}
        \tablefoot{%
            The models are the Navarro--Frenk--White~(NFW), Hernquist--Zhao~(HZ), and broken power-law~(BPL) profiles.
            For each model the natural logarithm of the Bayesian evidence and the decimal logarithm of the Bayes factor are shown.%
        }
    \end{table}
    The use of Bayesian evidence ensures that the different models employed with the same method can be fairly compared, taking into account that these models have different degrees of freedom.
    We assume the prior probabilities of the models are equal.
    Models are compared by taking the ratio of their Bayesian evidence~$Z$ or equivalently the difference of their $\log_{10}(Z)$, with the model with the largest $Z$ being favoured.
    The ratios or differences are interpreted using a scale; we will use the scale of \citet[their Appendix~B]{Jeffreys-1961-TP-C-3}.
    According to this scale, a ratio of $100$ or $\Delta\log_{10}(Z) = 2$ is required for a decisive result.
    It is not possible to compare a model from one table to one from the other table because of the differences in the method of CJAM and pyGravSphere.

    In all cases, the differences between the models are small.
    Among the CJAM models, the FDM profile has the largest Bayesian evidence.
    The Bayes factors indicate that the preference of FDM over SIDM is strong, but by no means significant, while FDM is only barely preferred over CDM.
        The preference for CDM over SIDM is substantial.
    It is therefore not possible to rule out any of the three dark-matter theories with the current data.
    For the pyGravSphere models, the broken power-law model is substantially preferred over the NFW model and the Hernquist--Zhao model.
    The modest strength of the evidence for the broken power-law model indicates that moving away from an NFW-like profile with a logarithmic slope of $-3$ at large radii is not required at present.
    Thus we find no conclusive evidence for tidal stripping or truncation at the probed radii.
    Further data at larger radii will help constrain the effect of tidal stripping.

\section{Discussion}
\label{sec:discussion}
    The mass--concentration relation between the virial mass~$M_{200}$ and the concentration parameter $c_{200} \coloneqq r_{200}/r_\mathrm{s}$ of \citet{Dutton-2014-MNRAS-441-3359} predicts $\log_{10} c_{200} \approx 1.3$ for an NFW halo with a virial mass equal to that of Eri~2 at redshift zero, with a scatter of $0.11\,\mathrm{dex}$, but was calibrated on a simulation with significantly higher virial masses ($M_{200} \gtrsim 10^{12}\,h^{-1}\,M_\odot$).
    Using the semi-analytical relation of COMMAH~\citep{Correa-2015-MNRAS-450-1514, Correa-2015-MNRAS-450-1521, Correa-2015-MNRAS-452-1217} we calculate a predicted concentration $\log_{10} c_{200} \approx 1.2$.
    Our determinations of $\log_{10} c_{200}$ for the CJAM models are more than one standard deviation higher.
    As $M_{200}$ and $c_{200}$ are among our less well constrained parameters, we also perform a comparison in the space of two better constrained parameters for the CDM/NFW profile of CJAM.
    Given the recovered $r_\mathrm{s}$, we predict the density at $100\,\mathrm{pc}$ assuming the \citet{Dutton-2014-MNRAS-441-3359} mass--concentration relation, $\rho(100\,\mathrm{pc}) = 10^{7.95^{+0.34}_{-0.59}}\,M_\sun\,\mathrm{kpc}^{-3}$.
    Compared to the recovered $\rho_2 = 10^{8.92^{+0.29}_{-0.26}}\,M_\sun\,\mathrm{kpc}^{-3}$, this prediction is over two combined standard deviations lower, indicating the tension between $M_{200}$ and $c_{200}$ is even larger than suggested at face value.
    Satellite dwarf galaxies are biased towards larger concentrations because higher-concentration dwarf galaxies are more likely to survive accretion by a Milky Way--mass galaxy~\citep{Nadler-2018-ApJ-859-129}.
    This bias might explain (part of) the tension we see.

    Using the stellar mass–to–halo mass relation of \citet{Behroozi-2013-ApJ-770-57} with the stellar mass-to-light ratio of $1.56$ derived in \citetalias{Zoutendijk-2020-A&A-635-A107}, we expect a virial mass-to-light ratio $\Upsilon_{200} \approx 10^{2.9}\,M_\sun\,L_\sun^{-1}$ for Eri~2.
    Most of our models agree with this value, but there is a substantial tension for the SIDM and broken power-law models.
    Our half-light mass-to-light ratios $\Upsilon_{1/2}$ are all consistent with the value $420^{+210}_{-140}\,M_\sun\,L_\sun^{-1}$ found by \citet{Li-2017-ApJ-838-8}.

    The values that we find for the astrophysical factors are typical for dwarf spheroidal and ultra-faint dwarf galaxies~\citep{Bonnivard-2015-MNRAS-453-849, 2020arXiv200201229A}.
    Eri~2 is therefore not the most interesting single target for observations concerning annihilation and decay signals, but it may be useful in a joint analysis of dwarf galaxies.
    \Citet{Bonnivard-2015-MNRAS-446-3002} have shown that the astrophysical factors can be biased by a factor of a few when an incorrect light profile model or halo triaxiality is assumed.
    We have assumed the light profile is exponential and the dark-matter halo is spherical, therefore this bias may be present.

    The self-interaction coefficient~$\Gamma$ can be described in terms of more conventional parameters by examining Equation~\eqref{eq:selfint} and considering that the mass change is $-2m$ per annihilation event, with $m$ being the mass of the dark-matter particle.
    Assuming a cross-section~$\sigma$ and a typical velocity~$v$, we derive
    \begin{equation}
        \Gamma = \frac{2\sigma v}{m}.
        \label{eq:xsec}
    \end{equation}
    Our constraints on the effective self-interaction coefficients therefore translate to $\sigma/m < 1.1 \times 10^{-36}\,(f/10)^{-1}(v/10\,\mathrm{km}\,\mathrm{s}^{-1})^{-1}\,\mathrm{cm}^2\,\mathrm{eV}^{-1}\,c^2$ at the 68-\% confidence level and $\sigma/m < 4.1 \times 10^{-36}\,(f/10)^{-1}(v/10\,\mathrm{km}\,\mathrm{s}^{-1})^{-1}\,\mathrm{cm}^2\,\mathrm{eV}^{-1}\,c^2$ at the 95-\% confidence level, where $f = 10$ and $v = 10\,\mathrm{km}\,\mathrm{s}^{-1}$ are of the right order of magnitude for ultra-faint dwarf galaxies.
    Much stronger constraints exist from combined observations of dwarf galaxies with the \emph{Fermi}/LAT and MAGIC gamma-ray telescopes~\citep{MAGIC-2016-JCAP-02-039}, equivalent to upper limits as low as ${\sim}10^{-43}\,(v/10\,\mathrm{km}\,\mathrm{s}^{-1})\,\mathrm{cm}^{2}\,\mathrm{eV}^{-1}\,c^2$.
    These constraints, however, are only valid for $10^1\,\mathrm{GeV}\,c^{-2} \leq m \leq 10^5\,\mathrm{GeV}\,c^{-2}$ and depend on the annihilation products, while our constraint is valid for all masses and annihilation products.
    The results from density profiles and gamma-ray searches are therefore complementary.

    \citet{Lin-2016-JCAP-03-009} remarked that $\Gamma$ can also represent self-interaction through scattering.
    Dark-matter particles can be scattered from the dense inner regions, where interactions are most likely, to the outer regions, where their contribution to the local density is negligible due to the much larger area.
    This is effectively equivalent to annihilation of dark-matter particles, but the strength of the effect depends on how frequent a scattering event leads to particles leaving the centre of the dark-matter halo.
    This frequency is currently unknown, therefore it is not possible to convert $\Gamma$ to a scattering cross section.
    Other profiles for SIDM exist that are designed specifically for a scattering self-interaction, such as the profiles of \citet{Kaplinghat-2014-PhRvL-113-021302} and \citet{Kaplinghat-2016-PhRvL-116-041302}, but these are outside the scope of this paper.
    \citet{2020arXiv200802529H} used the latter profile on 23~UFDs using literature kinematics and found no evidence for a non-zero self-interaction in these galaxies.

    Our lower limit on the FDM-particle mass of $m_\mathrm{a} > 4.0 \times 10^{-21}\,\mathrm{eV}\,c^{-2}$ at the 95-\% confidence level is incompatible with some results for other dwarf galaxies.
    \Citet{Chen-2017-MNRAS-468-1338} find $m_\mathrm{a} = 1.18^{+0.28}_{-0.24} \times 10^{-22}\,\mathrm{eV}\,c^{-2}$ or $m_\mathrm{a} = 1.79^{+0.35}_{-0.33} \times 10^{-22}\,\mathrm{eV}\,c^{-2}$, depending on the data set used, for the eight classical dwarf spheroidal galaxies.
    For the ultra-diffuse galaxy \object{Dragonfly~44}, \citet{Wasserman-2019-ApJ-885-155} find $m_\mathrm{a} = 3.3^{+10.3}_{-2.1} \times 10^{-22}\,\mathrm{eV}\,c^{_2}$.
    \Citet{Broadhurst-2020-PhRvD-101-083012} find $m_\mathrm{a} = 0.81^{+0.41}_{-0.21} \times 10^{-22}\,\mathrm{eV}\,c^{-2}$ for the ultra-diffuse galaxy \object{Antlia~II} and $m_\mathrm{a} = 1.07 \pm 0.08 \times 10^{-22}\,\mathrm{eV}\,c^{-2}$ when combined with four classical dwarf spheroidal galaxies.
    This discrepancy might indicate that the cores in the literature galaxies, which have higher masses than Eri~2, are formed by baryonic processes~\citep{Brooks-2014-ApJ-786-87, DiCintio-2014-MNRAS-437-415} and not (entirely) by FDM.
    Other constraints on FDM from Eri~2 have been derived from the survival of its star cluster~\citep{Marsh-2019-PhRvL-123-051103, ElZant-2020-MNRAS-492-877}.
    These constraints rule out at least the mass range between ${\sim}10^{-20}\,\mathrm{eV}\,c^{-2}$ and ${\sim}10^{-19}\,\mathrm{eV}\,c^{-2}$ and can likely be extended further, with some caveats.

    In simulations of spherically symmetric and relaxed FDM haloes a scaling relation between the size of the soliton, the mass of the FDM particle, and the virial mass of the halo is found \citep{Schive-2014-PhRvL-113-261302, 2020arXiv200701316N}:
    \begin{equation}
        r_\mathrm{c} = 1.6m_{22}^{-1} \Biggl(\frac{M_{200}}{10^9\,M_\sun}\Biggr)^{1/3}\,\mathrm{kpc}
    \end{equation}
    at redshift zero, where $r_\mathrm{c} = (9.1 \times 10^{-2})^{1/2} r_\mathrm{sol}$ and $m_{22} = m_\mathrm{a} / (10^{-22}\,\mathrm{eV}\,c^{-2})$.
    From the perspective of a single halo, $m_{22}r_\mathrm{c}$ is a constant.
    We find $m_{22}r_\mathrm{c} = 0.18^{+0.58}_{-0.30}\,\mathrm{kpc}$ directly from $m_\mathrm{a}$ and $r_\mathrm{sol}$, which is consistent with the expected $0.65^{+0.12}_{-0.17}\,\mathrm{kpc}$ based on the virial mass of Eri~2.

    \Citet{Amorisco-2017-ApJ-844-64} and \citet{Contenta-2018-MNRAS-476-3124} argue that the survival and projected location of the star cluster in Eri~2 imply that Eri~2 has a cored density profile.
    If the inner slope of the density profile is larger than ${\sim}0.2$--$0.25$, a cluster in a tight orbit would be tidally destroyed, while it would be unlikely to observe a cluster in a wide orbit so close in projection to the centre of Eri~2.
    The cluster could survive if it is stationary at the centre of the dark-matter halo of Eri~2, but that would mean that the photometric and gravitational centre of Eri~2 do not coincide.
    Our estimates of the inner slope are inconclusive in this respect: on the one hand, the broken power-law profile prefers a core, while on the other, the Hernquist--Zhao profile disagrees by nearly $2\sigma$.

    We have performed our pyGravSphere analysis with different numbers of stars per kinematic bin: 9 (the default of pyGravSphere for 92~stars in total), 11 (our fiducial analysis presented in this paper), 15, and 23.
    The recovered profiles for 11 and 15 stars per bin were consistent; we chose to use 11 stars per bin as it has more bins and could therefore potentially better capture the behaviour at small radii.
    The pyGravSphere profiles for 9 stars per bin had a much larger scale radius and lower characteristic density, inconsistent with both the 11 and 15 bin profiles and the CJAM profiles.
    Binning the stars by 23 yielded only two bins with a positive intrinsic velocity dispersion, which is too few for pyGravSphere to run.
    Therefore, as far as we can test, the profiles recovered by pyGravSphere seem stable with respect to the number of stars per bin, as long as a minimum number of stars per bin is met.
    We meet this requirement for our fiducial analysis with 11 stars per bin.

    Dynamical mass estimates are only correct if the system is in dynamical equilibrium.
    As we argued in \citetalias{Zoutendijk-2020-A&A-635-A107}, given that Eri~2 is currently close to its pericentre~\citep{Fritz-2018-A&A-619-A103} yet still $366\,\mathrm{kpc}$ removed from us~\citep{Crnojevic-2016-ApJL-824-L14}, it has not closely approached the Milky Way.
    Neither have any tidal features been detected in deep imaging~\citep{Crnojevic-2016-ApJL-824-L14}.
    Furthermore, the stars in Eri~2 are dominated by an old population~\citep{2020arXiv201200043S}.
    Therefore we do not expect a significant departure from dynamical equilibrium due to either tidal interactions with the Milky Way or stellar feedback.

    Another issue that can affect dynamical mass estimates is the presence of binary stars.
    Due to its orbital motion, the line-of-sight velocity of a binary star can change over time.
    Instead of the systemic velocity of the binary system, one sees another contribution on top of that, which may inflate measurements of velocity dispersion.
    We have observed our fields at multiple epochs for over a year.
    By combining the exposures before the data reduction, the velocity variation of short-period binary stars is blended into broadened spectral features.
    These should have the same centroid as the binary-systemic line-of-sight velocities and should therefore not impact our measurements.
    Longer-period binary systems typically have lower line-of-sight velocity deviations, so they are not expected to be a significant problem.
    Nevertheless, there remains much to be studied regarding the binary-star populations of UFDs.

    We have assumed that the dark-matter halo of Eri~2 is spherical, even though the stellar distribution is not.
    This could potentially bias the dark matter--density profiles.
    \citet{Read-2017-MNRAS-471-4541} have shown that GravSphere can become slightly biased for triaxial haloes, but the bias on the density profile is within the 95-\% confidence interval in most cases, and so is the mass within the half-light radius.
    This test was done with mock data resembling classical dwarf galaxies; as we have less data and larger measurement uncertainties, we expect any bias on the pyGravSphere density profiles due to triaxiality to be even smaller relative to the confidence intervals than for the mock classical dwarfs.
    As we obtain similar results with CJAM and pyGravSphere, the CJAM density profiles should also not be significantly biased.

    There is some uncertainty in the position of the centre of Eri~2.
    Mis-centring the spatial coordinates can affect the derived density profile, because the density measured at the centre of the coordinate system will be lower than the density at the true centre of the galaxy.
    This effect can lead to cored density profiles being measured for cuspy dark-matter haloes, or to core radii being biased to larger values for cored haloes.
    We do not detect a core or soliton for Eri~2 and provide upper limits for the core and soliton radii.
    Our upper limits on core and soliton radii could therefore be biased high, but this would strengthen rather than weaken the confidence level of these limits.

\section{Conclusions}
\label{sec:conclusions}
    We have presented new data from the MUSE-Faint survey of the ultra-faint dwarf galaxy Eridanus~2 ($M_\mathrm{V} = -7.1$, $M_* \approx 9 \times 10^4\, M_\sun$).
    Ultra-faint dwarf galaxies have the lowest baryonic fractions of all known galaxies; it is expected that the baryonic contents have not altered the dark matter--density profiles.
    We have modelled the dark matter--density profile of Eridanus~2 using stellar kinematics from MUSE-Faint and from the literature (92~stars in total) to constrain the properties of self-interacting and fuzzy dark matter and to compare these dark-matter models against each other and against cold dark matter.
    For modelling the density profiles we have used both CJAM and pyGravSphere, two codes that use different methods and assumptions, to test whether the recovery of the density profile is sensitive to the approach that is used.

    We constrained the core radius of the self-interacting dark-matter profile to $r_\mathrm{c} < 47\,\mathrm{pc}$ (68-\% confidence level) or $r_\mathrm{c} < 117\,\mathrm{pc}$ (95-\% confidence level).
    This translates into a constraint on the effective self-interaction coefficient: $f\Gamma < 2.2 \times 10^{-29}\,\mathrm{cm}^3\,\mathrm{s}^{-1}\,\mathrm{eV}^{-1}\,c^2$ (68-\% confidence level) or $f\Gamma < 8.1 \times 10^{-29}\,\mathrm{cm}^3\,\mathrm{s}^{-1}\,\mathrm{eV}^{-1}\,c^2$ (95-\% confidence level).
    These effective self-interaction coefficients are equivalent to the specific annihilation cross sections $\sigma/m < 1.1 \times 10^{-36}\,(f/10)^{-1}(v/10\,\mathrm{km}\,\mathrm{s}^{-1})^{-1}\,\mathrm{cm}^2\,\mathrm{eV}^{-1}\,c^2$ (68-\% confidence level) or $\sigma/m < 4.1 \times 10^{-36}\,(f/10)^{-1}(v/10\,\mathrm{km}\,\mathrm{s}^{-1})^{-1}\,\mathrm{cm}^2\,\mathrm{eV}^{-1}\,c^2$ (95-\% confidence level).
    These constraints apply for all dark matter--particle masses and are therefore complementary to the results from gamma-ray searches for annihilation signatures, which provide stronger constraints in a limited mass range.

    We constrained the soliton radius of the fuzzy dark-matter profile to $r_\mathrm{sol} < 7.2\,\mathrm{pc}$ (68-\% confidence level) or $r_\mathrm{sol} < 102\,\mathrm{pc}$ (95-\% confidence level).
    The equivalent constraint on the mass of the ultra-light dark-matter particle is $m_\mathrm{a} > 5.9 \times^{-20}\,\mathrm{eV}\,c^{-2}$ (68-\% confidence level) or $m_\mathrm{a} > 4.0 \times 10^{-21}\,\mathrm{eV}\,c^{-2}$ (95-\% confidence level).
    These constraints are inconsistent with particle masses for larger dwarf galaxies, which may indicate the cores in these larger dwarf galaxies are not caused by fuzzy dark matter.

    We could not consistently constrain the velocity anisotropy of Eridanus~2.
    CJAM and pyGravSphere prefer different values for the inner and outer slope of the density profile when these are free parameters of the profile, therefore we cannot draw conclusions about the survival or location of the star cluster.

    We found that CJAM and pyGravSphere recover similar dark matter--density profiles for Eridanus~2 when a cold dark matter~/ Navarro--Frenk--White profile is assumed in both cases.
    All profiles of CJAM and pyGravSphere are consistent within their uncertainties.
    The uncertainty on the profile and the difference between the profiles become larger near the centre of Eridanus~2, where the kinematic data are sparse.

    From the dark matter--density profiles we determined virial masses $M_{200} \sim 10^8\,M_\sun$, maximum circular velocities $V_\mathrm{max} \sim 10^{1.2}$--$10^{1.4}\,\mathrm{km}\,\mathrm{s}^{-1}$, half-light mass-to-light ratios $\Upsilon_{1/2} \sim 10^{2.5}\,M_\sun\,L_\sun^{-1}$, and astrophysical factors $J(\alpha_\mathrm{c}^J) \sim 10^{11}\,M_\sun^2\,\mathrm{kpc}^{-5}$ and $D(\alpha_\mathrm{c}^D) \sim 10^2$--$10^{2.5}\,M_\sun\,\mathrm{kpc}^{-2}$.
    The half-light mass-to-light ratio is consistent with the literature and the astrophysical factors are typical for dwarf galaxies.
    For CJAM with the cold dark-matter model, the values are $M_{200} = 10^{7.89^{+0.60}_{-0.36}}\,M_\sun$, $V_\mathrm{max} = 10^{1.19^{+0.09}_{-0.06}}\,\mathrm{km}\,\mathrm{s}^{-1}$, $\Upsilon_{1/2} = 10^{2.59^{+0.10}_{-0.11}}\,M_\sun\,L_\sun^{-1}$, $J(\alpha_\mathrm{c}^J) = 10^{10.94^{+0.57}_{-0.38}}\,M_\sun^2\,\mathrm{kpc}^{-5}$, and $D(\alpha_\mathrm{c}^D) = 10^{2.31^{+0.34}_{-0.22}}\,M_\sun\,\mathrm{kpc}^{-2}$.
    The concentration $c \sim 10^{1.5}$--$10^2$ ($c = 10^{1.95^{+0.40}_{-0.39}}$ for CJAM with cold dark matter) is for several profiles higher than the expected value for a galaxy of this virial mass, but this may be because Eridanus~2 is a satellite of the Milky Way.

    We found a weak preference for fuzzy dark matter over cold dark matter and substantial evidence for cold dark matter over self-interacting dark matter.
    The evidence to prefer fuzzy dark matter over self-interacting dark matter is strong.
    This indicates a preference for a cusp above a core, but also for a soliton above a cusp.
    None of the models are preferred decisively over any other, therefore it is not possible to rule out cold dark matter, self-interacting dark matter, or fuzzy dark matter.

    With MUSE-Faint we have been able to significantly increase the number of stars with spectroscopy inside the half-light radius of Eridanus~2 and have extended the available data to smaller radii.
    Nevertheless, it remains challenging to obtain a large sample of stellar line-of-sight velocities in such a faint and far-away system.
    Improvements of the constraints on the inner dark matter--density profile of Eridanus~2 and its implications for the nature and properties of dark matter would require deeper observations or observations at a higher spectral resolution.
    Deeper observations could improve the line-of-sight velocity measurements and could provide access to fainter stars, but would be a costly undertaking.
    A higher spectral resolution could significantly decrease the velocity uncertainties, but current high-resolution spectrographs are not able to reach the spatial resolution required for these crowded systems.
    It would also be valuable to extend the current study to multiple ultra-faint dwarf galaxies and test whether our conclusions also hold for other systems.

    \begin{acknowledgements}
        We thank the anonymous referee for their helpful comments, which improved the manuscript.
        SLZ wishes to thank Anna Genina and Justin~I. Read for interesting and useful discussions, and Mariana~P. J{\'u}lio for asking helpful questions.
        SLZ acknowledges support by The Netherlands Organisation for Scientific Research~(NWO) through a TOP Grant Module~1 under project number 614.001.652.
        JB acnowledges support by Funda\c{c}\~{a}o para a Ci\^{e}ncia e a Tecnologia~(FCT) through the research grants UID/FIS/04434/2019, UIDB/04434/2020, UIDP/04434/2020 and through the Investigador FCT Contract No.\ IF/01654/2014/CP1215/CT0003.

        Based on observations made with ESO Telescopes at the La Silla Paranal Observatory under programme IDs 0100.D-0807, 0102.D-0372, 0103.D-0705, and 0104.D-0199.

        This research has made use of Astropy~\citep{Robitaille-2013-A&A-558-A33, AstropyCollaboration-2018-AJ-156-123}, corner.py~\citep{ForemanMackey-2016-JOSS-1-24}, matplotlib~\citep{Hunter-2007-CSE-9-90}, NASA's Astrophysics Data System Bibliographic Services, NumPy~\citep{Harris-2020-Natur-585-357}, SciPy~\citep{2020NatMe..17..261V}, and the colour schemes of \citet{tolcolor}.

        This work has made use of data from the European Space Agency (ESA) mission
        {\it Gaia} (\url{https://www.cosmos.esa.int/gaia}), processed by the {\it Gaia}
        Data Processing and Analysis Consortium (DPAC,
        \url{https://www.cosmos.esa.int/web/gaia/dpac/consortium}). Funding for the DPAC
        has been provided by national institutions, in particular the institutions
        participating in the {\it Gaia} Multilateral Agreement.

        This research made use of Montage.
        It is funded by the National Science Foundation under Grant Number ACI-1440620, and was previously funded by the National Aeronautics and Space Administration's Earth Science Technology Office, Computation Technologies Project, under Cooperative Agreement Number NCC5-626 between NASA and the California Institute of Technology.
    \end{acknowledgements}

    \bibliographystyle{aa}
    \bibliography{Zoutendijk_Eridanus2-profile}

    \appendix
    \onecolumn
\section{Table of kinematics}
    In Table~\ref{tab:selection} we list the positions and line-of-sight velocities of the stars used for the kinematic analysis in this paper.
    \begin{longtable}{cccc}
        \caption{%
            Final selection of stars in Eridanus~2 for the kinematic analysis.%
            \tablefoot{%
                The columns are the source ID, the right ascension and the declination in degrees, and the line-of-sight velocity and its measurement uncertainty in $\mathrm{km}\,\mathrm{s}^{-1}$.
                The source IDs below $2\,000\,000$ are consistent with those in \citetalias{Zoutendijk-2020-A&A-635-A107}.
                Source IDs starting with $2\,000\,000$ are sources from \citet{Li-2017-ApJ-838-8} that have no counterpart in the source extraction catalogue of \citetalias{Zoutendijk-2020-A&A-635-A107}.
                The right ascension and declination have been calibrated to \emph{Gaia} Data Release~2~\citep{GaiaCollaboration-2016-A&A-595-A1, GaiaCollaboration-2018-A&A-616-A1, Lindegren-2018-A&A-616-A2}.%
            }
        }
        \label{tab:selection}\\
        \hline
        \hline
        ID        & RA (deg)   & Dec. (deg)  & LOS velocity ($\mathrm{km}\,\mathrm{s}^{-1}$) \\
        \hline
        \endfirsthead
        \caption{Continued.}\\
        \hline
        \hline
        ID        & RA (deg)   & Dec. (deg)  & LOS velocity ($\mathrm{km}\,\mathrm{s}^{-1}$) \\
        \hline
        \endhead
        \hline
        \endfoot
        \hline
        \endlastfoot
           $1058$ & $56.06437$ & $-43.53266$ &  $72.3 \pm 20.3$ \\
           $2348$ & $56.06852$ & $-43.52907$ &  $64.1 \pm 14.7$ \\
           $3932$ & $56.08301$ & $-43.54452$ &  $67.5 \pm 21.0$ \\
           $4448$ & $56.07485$ & $-43.52340$ &  $61.1 \pm 21.9$ \\
           $4630$ & $56.08690$ & $-43.54593$ &  $53.9 \pm 16.4$ \\
           $4866$ & $56.08551$ & $-43.54109$ & $100.3 \pm 12.3$ \\
           $5256$ & $56.08346$ & $-43.53403$ &  $73.2 \pm  9.4$ \\
           $6227$ & $56.08961$ & $-43.53808$ &  $54.0 \pm 22.4$ \\
           $6664$ & $56.08621$ & $-43.52849$ &  $45.1 \pm 13.2$ \\
           $9242$ & $56.09260$ & $-43.52854$ &  $79.2 \pm 23.8$ \\
           $9304$ & $56.09734$ & $-43.53763$ &  $47.9 \pm 14.2$ \\
           $9653$ & $56.09155$ & $-43.52363$ &  $61.1 \pm 25.8$ \\
          $11171$ & $56.09772$ & $-43.52349$ &  $65.7 \pm 20.5$ \\
          $11935$ & $56.10766$ & $-43.53628$ &  $96.0 \pm 18.2$ \\
          $12933$ & $56.11073$ & $-43.53324$ &  $59.2 \pm 27.9$ \\
          $13257$ & $56.11460$ & $-43.53779$ &  $75.8 \pm 25.2$ \\
          $13549$ & $56.11108$ & $-43.52766$ &  $85.4 \pm 12.7$ \\
          $14541$ & $56.11801$ & $-43.53130$ &  $83.9 \pm 24.0$ \\
          $14551$ & $56.12156$ & $-43.53821$ &  $47.0 \pm 13.2$ \\
          $14927$ & $56.12031$ & $-43.53184$ &  $73.3 \pm 18.3$ \\
        $1002926$ & $56.06117$ & $-43.52640$ &  $73.5 \pm  1.2$ \\
        $1003016$ & $56.06721$ & $-43.53447$ &  $68.5 \pm  8.2$ \\
        $1003965$ & $56.07701$ & $-43.55105$ &  $83.1 \pm  3.0$ \\
        $1004032$ & $56.07494$ & $-43.54397$ &  $78.3 \pm  7.8$ \\
        $1004756$ & $56.07025$ & $-43.53160$ &  $88.6 \pm  8.8$ \\
        $1005369$ & $56.06965$ & $-43.52886$ &  $79.2 \pm  5.1$ \\
        $1005680$ & $56.07367$ & $-43.53681$ &  $85.7 \pm  6.8$ \\
        $1006056$ & $56.08153$ & $-43.55039$ &  $59.3 \pm  3.5$ \\
        $1006522$ & $56.07226$ & $-43.52913$ &  $90.9 \pm 12.5$ \\
        $1007072$ & $56.07965$ & $-43.54091$ &  $86.1 \pm  9.3$ \\
        $1007081$ & $56.08013$ & $-43.54021$ &  $98.2 \pm 13.3$ \\
        $1007232$ & $56.08398$ & $-43.54946$ &  $75.7 \pm  4.7$ \\
        $1007817$ & $56.08566$ & $-43.54801$ &  $91.5 \pm  7.7$ \\
        $1007943$ & $56.08618$ & $-43.55232$ &  $84.5 \pm  1.5$ \\
        $1008083$ & $56.07548$ & $-43.52653$ &  $78.3 \pm  5.6$ \\
        $1008946$ & $56.07992$ & $-43.53195$ &  $94.8 \pm  5.1$ \\
        $1009001$ & $56.07691$ & $-43.52592$ &  $76.3 \pm  4.3$ \\
        $1009750$ & $56.07605$ & $-43.51971$ &  $63.6 \pm 11.6$ \\
        $1010022$ & $56.07599$ & $-43.52005$ &  $79.2 \pm  1.4$ \\
        $1010255$ & $56.07916$ & $-43.52591$ &  $86.0 \pm  3.7$ \\
        $1010560$ & $56.08680$ & $-43.54108$ &  $86.2 \pm  1.0$ \\
        $1010966$ & $56.08882$ & $-43.54120$ &  $62.3 \pm  9.5$ \\
        $1010988$ & $56.08438$ & $-43.53642$ &  $79.9 \pm  0.9$ \\
        $1011039$ & $56.08312$ & $-43.52889$ &  $69.4 \pm  4.9$ \\
        $1012006$ & $56.09504$ & $-43.54725$ &  $56.9 \pm  6.4$ \\
        $1012321$ & $56.09129$ & $-43.53950$ & $100.8 \pm  8.0$ \\
        $1013259$ & $56.08689$ & $-43.52648$ &  $74.7 \pm 14.9$ \\
        $1013271$ & $56.09513$ & $-43.54466$ &  $95.8 \pm  3.5$ \\
        $1013803$ & $56.08677$ & $-43.52694$ &  $88.6 \pm  7.1$ \\
        $1014555$ & $56.09416$ & $-43.53802$ &  $77.8 \pm  9.7$ \\
        $1017156$ & $56.09828$ & $-43.53589$ &  $74.1 \pm 10.2$ \\
        $1017445$ & $56.09230$ & $-43.52332$ &  $81.1 \pm  6.2$ \\
        $1018571$ & $56.09774$ & $-43.52937$ &  $75.8 \pm  3.8$ \\
        $1018845$ & $56.09547$ & $-43.52370$ &  $71.9 \pm  7.9$ \\
        $1019322$ & $56.09626$ & $-43.52348$ &  $74.5 \pm  1.2$ \\
        $1019765$ & $56.10630$ & $-43.53993$ &  $76.0 \pm  5.6$ \\
        $1019801$ & $56.10251$ & $-43.53367$ &  $97.3 \pm  6.4$ \\
        $1021252$ & $56.10830$ & $-43.53636$ &  $82.9 \pm  4.0$ \\
        $1021910$ & $56.10425$ & $-43.52568$ & $100.7 \pm  8.9$ \\
        $1022334$ & $56.10586$ & $-43.52943$ &  $71.4 \pm  3.6$ \\
        $1022351$ & $56.11369$ & $-43.54099$ &  $57.1 \pm 10.1$ \\
        $1022417$ & $56.10670$ & $-43.52748$ &  $80.2 \pm  9.2$ \\
        $1023228$ & $56.11503$ & $-43.53988$ &  $82.3 \pm  5.2$ \\
        $1024420$ & $56.11196$ & $-43.52721$ &  $73.3 \pm  4.3$ \\
        $1025752$ & $56.11125$ & $-43.52308$ &  $78.1 \pm  0.9$ \\
        $1026606$ & $56.11505$ & $-43.52693$ &  $79.6 \pm  1.2$ \\
        $1026881$ & $56.12141$ & $-43.53211$ &  $63.1 \pm  8.1$ \\
        $1027080$ & $56.12490$ & $-43.54157$ &  $84.7 \pm  1.9$ \\
        $1027101$ & $56.12301$ & $-43.53626$ &  $80.0 \pm  6.3$ \\
        $1027929$ & $56.11814$ & $-43.52022$ &  $83.9 \pm  5.0$ \\
        $1027958$ & $56.11869$ & $-43.52139$ &  $68.0 \pm  5.6$ \\
        $1030234$ & $56.12404$ & $-43.53005$ &  $71.9 \pm  0.9$ \\
        $2000001$ & $56.00955$ & $-43.53305$ &  $69.8 \pm  1.6$ \\
        $2000002$ & $56.02915$ & $-43.52877$ &  $77.9 \pm  1.0$ \\
        $2000003$ & $56.04649$ & $-43.51453$ &  $65.4 \pm  2.3$ \\
        $2000004$ & $56.05139$ & $-43.51837$ &  $75.1 \pm  2.6$ \\
        $2000005$ & $56.05287$ & $-43.50876$ &  $91.2 \pm  1.5$ \\
        $2000007$ & $56.06543$ & $-43.50896$ &  $65.8 \pm  1.6$ \\
        $2000008$ & $56.06747$ & $-43.54544$ &  $74.0 \pm  0.8$ \\
        $2000010$ & $56.08023$ & $-43.50531$ &  $81.7 \pm  3.0$ \\
        $2000014$ & $56.08915$ & $-43.50587$ &  $77.3 \pm  1.1$ \\
        $2000016$ & $56.10013$ & $-43.54549$ &  $67.7 \pm  0.9$ \\
        $2000017$ & $56.11077$ & $-43.54558$ &  $69.5 \pm  1.1$ \\
        $2000019$ & $56.11480$ & $-43.54807$ &  $75.1 \pm  2.4$ \\
        $2000020$ & $56.11801$ & $-43.54748$ &  $71.8 \pm  0.8$ \\
        $2000021$ & $56.12240$ & $-43.52515$ &  $74.7 \pm  1.2$ \\
        $2000023$ & $56.12624$ & $-43.51339$ &  $79.3 \pm  2.4$ \\
        $2000024$ & $56.12985$ & $-43.55450$ &  $89.6 \pm  1.3$ \\
        $2000025$ & $56.13921$ & $-43.55537$ &  $66.7 \pm  2.0$ \\
        $2000026$ & $56.16179$ & $-43.50427$ &  $74.2 \pm  1.8$ \\
        $2000027$ & $56.16557$ & $-43.51079$ &  $68.4 \pm  2.2$ \\
        $2000028$ & $56.19012$ & $-43.49878$ &  $80.5 \pm  1.0$ \\
    \end{longtable}

    \twocolumn
\section{Multi-Gaussian expansions of the density profiles}
\label{app:mge}
    The JAM method uses MGEs of the density profiles to speed up its calculations.
    The MGE itself is expensive, but for simple profiles it has to be computed only once.
    Some of our profiles, however, have parameters that modify their shape, which necessitates a different MGE for each combination of values of these parameters.
    Because fitting the MGE separately for every Monte Carlo sample would be prohibitively expensive, we instead interpolate between MGEs fitted at a limited number of points in the parameter space.
    In this appendix we describe the fitting and interpolation procedures to obtain MGEs of our profiles.

    For a scale-invariant profile like the NFW profile, the MGE can be done only once, and the amplitude and standard deviations of the Gaussians can be rescaled to fit the NFW profile for any combination of characteristic density and scale radius.
    We perform the MGE on the NFW profile with 16~Gaussians, by least-squares fitting at 64~logarithmically spaced points from $10^{-3}\,r_\mathrm{s}$ to $10^2\,r_\mathrm{s}$, weighting the residuals with the value of the NFW profile.
    The range of points chosen for fitting is deliberately broad, in order to be sure that the observations are contained within the limits of this range for any reasonable choice of $r_\mathrm{s}$.
    In addition, the projection of the profile on the sky has as a result that any radius larger than the projected radius is observed along the line of sight, and therefore the MGE needs to also reproduce the profile at radii larger than the largest projected radius.
    The resulting MGE is an accurate reproduction over the fitting range with deviations on the order of at most $1\,\%$ in both density and cumulative mass.

    The distribution of tracers is fit with the same number of Gaussians to the exponential profile found by \citet{Crnojevic-2016-ApJL-824-L14}, which has an effective radius of $2.31\,\textrm{arcmin}$ and a central surface brightness of $27.2\,\mathrm{mag}$.
    We also adopt their position angle of $72.6\,\deg$ and use their ellipticity $\epsilon = 0.48$ to calculate a flattening of $q = 1 - \epsilon = 0.52$.
    To reduce the computational complexity we fix the inclination to the default value of $90^\circ$, corresponding to an edge-on system.
    The exponential drop-off is hard to reproduce over large orders of magnitude in radius, so we limit the fit to the range of $10^{-3}\,R_\mathrm{e}$ to $10\,R_\mathrm{e}$.
    A smaller range suffices here as we know the value of $R_\mathrm{e}$ from photometry.
    The accuracy of the MGE is sub-percent over the fitting range for the density and for at least an extra magnitude of larger radii for the cumulative mass.
    The resulting fit is valid for all dark-matter models and does not need to be rescaled for different dark-matter parameters.

    Unfortunately, a single MGE is not possible for the SIDM and FDM profiles.
    The computational expense of redoing the MGE for each combination of profile parameters is prohibitively large, therefore we need to approximate the MGE with a faster method.
    For the SIDM profile, we perform the same procedure as for the NFW profile for 101~logarithmically spaced values of the core radius, from $10^{-2}\,r_\mathrm{s}$ to $r_\mathrm{s}$.
    We exclude models with a core radius larger than the NFW scale radius, because such large cores are not expected giving existing work on dark matter--density profiles in dwarf galaxies (see e.g.\ the high-resolution simulations analysed by \citealt{2020arXiv200410817L}).
    The amplitudes and standard deviations of the 16~Gaussians vary smoothly with the core size, so we interpolate over these 101~results, supplemented with the NFW profile corresponding to a core size of zero, with quadratic splines.
    The 16~Gaussians returned by supplying the interpolator with a core radius can be rescaled with the characteristic density and scale radius, as for the NFW profile.
    We find that the resulting interpolated MGE is sufficiently close to a real MGE: the deviation from the original profile is still less than one percent.  The interpolated MGE is also sufficiently fast for our purposes.

    The FDM profile has the largest number of parameters of all our models and is therefore the most complex to expand into Gaussians.
    In addition to this, the profile proposed by \citet{Marsh-2015-MNRAS-451-2479} has a sharp transition from the soliton part to the NFW part, which is very hard to approximate with a sum of Gaussians.
    Since this sharp transition was assumed for sake of simplicity and the lack of detailed knowledge about the true transition, we find it justified to make a different simplifying assumption that suits our needs better.
    We approximate the FDM profile with the sum of a soliton and a cored NFW profile, the latter being the same profile that we use to model SIDM:
    \begin{align}
        \begin{split}
            &\tilde{\rho}_\mathrm{FDM}(r; \tilde{\rho}_{\mathrm{sol},0}, r_\mathrm{sol}, \rho_0, r_\mathrm{c}, r_\mathrm{s}) =\\
            &\qquad \rho_{\mathrm{sol}}(r; \tilde{\rho}_{\mathrm{sol},0}, r_\mathrm{sol}) + \rho_\mathrm{SIDM}(r; \rho_0, r_\mathrm{c}, r_\mathrm{s}),
        \end{split}
        \label{eq:pfdm}
    \end{align}
    with $\tilde{\rho}_{\mathrm{sol},0} = \rho_{\mathrm{sol},0} - \rho_\mathrm{SIDM}(0; \rho_0, r_\mathrm{c}, r_\mathrm{s})$ to ensure the characteristic density is correct.
    We find this is a good approximation of $\rho_\mathrm{FDM}$ for a certain value of~$r_\mathrm{c}$, depending on the values of the other parameters.
    Minimizing the difference between the two profiles at 64~logarithmically spaced points from $10^{-3}\,r_\mathrm{sol}$ to $10^2\,r_\mathrm{sol}$ while varying $r_\mathrm{sol}/r_\mathrm{s}$ over 101~logarithmically spaced values between $10^{-2}$ and $1$, and $\varepsilon$ over 101~logarithmically spaced values between $10^{-5}$ and $1/2$, yields values of $r_\mathrm{c}$ roughly following the relation
    \begin{equation}
        r_\mathrm{c} = C_0 r_\mathrm{sol} \varepsilon^{C_1},
        \label{eq:pfdmrc}
    \end{equation}
    where $C_0$ and $C_1$ are constants.
    Enforcing this relation and repeating the minimization gives $C_0 \approx 0.281$ and $C_1 \approx -0.0923$.
    By construction the profiles are identical at the centre and towards infinity.
    The largest deviation is at the transition radius, but is very localized, and considering the original profile is only an approximation at this point as well, this is not a problem.
    We then approximate the soliton profile with 16~Gaussians by fitting to 64~logarithmically spaced points from $10^{-3}\,r_\mathrm{sol}$ to $2r_\mathrm{sol}$, at which point the soliton density has declined so far that it is negligible compared to the NFW part of the FDM profile.
    The soliton and SIDM fits can be individually rescaled depending on the parameters of the profile, and are then appended to form a MGE for the FDM profile.
    With this result we can approximate the original $\rho_\mathrm{FDM}$ with the MGE of $\tilde{\rho}_\mathrm{FDM}$.
    At the transition radius the deviation can be very large, but elsewhere the accuracy is on the level of a few percent's deviation from the original density profile.

\section{Unbiased estimators of intrinsic velocity moments}
\label{app:vmom}
    With a small number of stars and relatively high uncertainties on their velocities, careful estimation of the observed velocity moments and the correction term for measurement uncertainties is needed for a reliable dynamical analysis of ultra-faint dwarf galaxies.
    In this appendix we derive unbiased estimators of the second and fourth intrinsic velocity moments as well as an estimator for the uncertainty on the second intrinsic velocity moment.
    The estimators are exact when each velocity measurement has the same measurement uncertainty and are approximations when the measurement uncertainties are different.

    Suppose we have $N$ velocity measurements $v_1, v_2, \ldots, v_N$ with measurement uncertainties $\varepsilon_1, \varepsilon_2, \ldots, \varepsilon_N$.
    The velocities can be divided into bins with $n$ measurements: $v_{j+1}, v_{j+2}, \ldots, v_{j+n}$.
    We assume the intrinsic --~as opposed to the observed~-- velocities in each bin are drawn from the same distribution, and that the distributions of each bin have the same mean: the systemic velocity.
    Furthermore, we also assume the measurement errors are normally distributed around zero with a standard deviation equal to the measurement uncertainty.
    We are prevented from straightforwardly calculating the moments of the observed velocities by two effects: the measurement errors inflate the observed velocity moments, leading to a difference between the intrinsic and observed distributions, and the sample moments are biased estimators of the true moments of the observed distributions.

    In the case of equal uncertainties on all measured velocities, the unbiased estimators of the intrinsic velocity moments can be determined exactly.
    For now we will further assume that all measurement uncertainties are equal to $\varepsilon$.
    We begin by calculating the sample mean
    \begin{equation}
        m = \frac{1}{N} \sum_{i=1}^N v_i
        \label{eq:smean}
    \end{equation}
    of all measurements and the $r$th sample central moments
    \begin{equation}
        m_r = \frac{1}{n} \sum_{i=j+1}^{j+n} (v_i-m)^r
        \label{eq:scmom}
    \end{equation}
    in each bin.
    The correction for the inflation of the moments by the measurement errors can be done using the cumulants, because cumulants have the property
    \begin{equation}
        \kappa_r(X + Y) = \kappa_r(X) + \kappa_r(Y),
        \label{eq:cumul}
    \end{equation}
    for random variables $X$ and $Y$; in this case the intrinsic velocities and the measurement errors.
    The symmetrically unbiased estimators of the cumulants of a distribution are the $k$~statistics~\citep{Fisher-1930-ProcLondMathSoc-s2_30-199}, which for the second and fourth cumulants are
    \begin{align}
        k_2 &= \frac{n}{n-1} m_2,
        \label{eq:kstat2}\\
        k_4 &= \frac{n^2 \Big[(n+1)m_4 - 3(n-1)m_2^2\Big]}{(n-1)(n-2)(n-3)}.
        \label{eq:kstat4}
    \end{align}
    The second and fourth cumulants of a normal distribution~$\mathcal{N}(\mu, \sigma^2)$ are $\kappa_2 = \sigma^2$ and $\kappa_4 = 0$.
    The distribution $\mathcal{N}(0, \varepsilon^2)$ of measurement errors therefore has cumulants $\kappa_{2,\text{err}} = \varepsilon^2$ and $\kappa_{4,\text{err}} = 0$.
    Correcting for the measurement uncertainty, using the properties of cumulants, the estimators of the intrinsic second and fourth cumulants are
    \begin{align}
        k_{2,\text{int}} &= k_2 - \kappa_{2,\text{err}},
        \label{eq:k2int}\\
        k_{4,\text{int}} &= k_4 - \kappa_{4,\text{err}}.
        \label{eq:k4int}
    \end{align}
    Using equations \eqref{eq:kstat2} and \eqref{eq:kstat4}, this can be converted to the second and fourth intrinsic sample central moments:
    \begin{align}
        m_{2,\text{int}} &= \frac{n-1}{n} k_{2,\text{int}},
        \label{eq:m2int}\\
        m_{4,\text{int}} &= \frac{\frac{(n-1)(n-2)(n-3)}{n^2}k_{4,\text{int}} + 3(n-1)m_{2,\text{int}}^2}{n+1}.
        \label{eq:m4int}
    \end{align}
    The symmetrically unbiased estimators of the central moments of a distribution are given by the $h$~statistics~\citep{Dwyer-1937-AnnMathStat-8-21}.
    We can therefore estimate the intrinsic central moments in each bin with
    \begin{align}
        h_{2,\text{int}} &= \frac{n}{n-1} m_{2,\text{int}} = k_{2,\text{int}},
        \label{eq:h2int}\\
        h_{4,\text{int}} &= \frac{n\Big[(n^2-2n+3)m_{4,\text{int}} - 3(2n-3)m_{2,\text{int}}^2\Big]}{(n-1)(n-2)(n-3)}.
        \label{eq:h4int}
    \end{align}

    In our case, where each velocity~$v_i$ has its own uncertainty~$\varepsilon_i$, there is no exact solution.
    Like \citet{vandeVen-2006-A&A-445-513}, we can try to approximate the correction of the cumulants with a single value.
    If the individual errors are interpreted as being drawn from a single distribution, the expected values of the second and fourth moments of this distribution are the averages of the same moments of the individual distributions:
    We suppose the measurement errors are drawn from a single distribution and want to find the cumulants of this distribution.
    For an infinite number of draws, the sample raw moments converge to the true raw moments of a distribution:
    \begin{equation}
        \mu'_r(X) = \lim_{N\to\infty} \frac{1}{N} \sum_{i=1}^N x_i^r.
        \label{eq:infmom}
    \end{equation}
    If the draws from this supposed single distribution are to be equivalent to draws from $n$ separate distributions of measurement errors, we can group draws from the same distribution together and write the above summation as
    \begin{equation}
        \mu'_r(X) = \frac{1}{n} \sum_{i=0}^{n-1} \lim_{N/n\to\infty} \frac{1}{N/n} \sum_{j=1}^{N/n} x_{ni+j}^r.
        \label{eq:infmomsplit}
    \end{equation}
    Comparing equations~\eqref{eq:infmom} and \eqref{eq:infmomsplit}, we can see that the latter equation is the average of the moments of the individual distributions.
    The central moments of the single distribution must therefore be
    \begin{align}
        \tilde{\mu}_{2,\text{err}} &= \tilde{\mu}'_{2,\text{err}} = \frac{1}{n} \sum_{i=j+1}^{j+n} \varepsilon_i^2,
        \label{eq:m2err}\\
        \tilde{\mu}_{4,\text{err}} &= \tilde{\mu}'_{4,\text{err}} = \frac{1}{n} \sum_{i=j+1}^{j+n} 3\varepsilon_i^4,
        \label{eq:m4err}
    \end{align}
    where we have used the assumption that the individual error distributions are normal distributions $\mathcal{N}(0, \varepsilon_i^2)$ centred around zero, making the raw moments equal to the central moments.
    The cumulants to use as approximate correction terms in equations \eqref{eq:k2int} and \eqref{eq:k4int} are therefore by definition
    \begin{align}
        \tilde{\kappa}_{2,\text{err}} &= \tilde{\mu}_{2,\text{err}},
        \label{eq:k2err}\\
        \tilde{\kappa}_{4,\text{err}} &= \tilde{\mu}_{4,\text{err}} - 3\tilde{\mu}_{2,\text{err}}^2.
        \label{eq:k4err}
    \end{align}
    The correction for the second cumulant is the same as used by \citet{vandeVen-2006-A&A-445-513}.

    There are two sources of uncertainty on the intrinsic moment estimators: measurement uncertainties and finite sampling.
    Both of these are reflected in the variance of the moments of the measured distribution.
    Remembering that the variance is the second raw moment of an estimator~$f_r$, we can write
    \begin{equation}
        \begin{split}
            \mu_2(f_r) &= \mathrm{E}\Bigl[(f_r - \mathrm{E}[f_r])^2\Bigr]\\
            &= \mathrm{E}\Bigl[(f_r)^2\Bigr] - (\mathrm{E}[f_r])^2 = \mu'_2(f_r) - {\mu'_1}^2(f_r),
        \end{split}
    \end{equation}
    where $\mathrm{E}$ denotes the expectation value.
    From the equations and tables of \citet{Dwyer-1937-AnnMathStat-8-21} it follows that the variance of the second sample central moment is
    \begin{equation}
        \mu_2(m_2) = \frac{(n-1)\Bigl((n-1)\mu_4-(n-3)\mu_2^2\Bigr)}{n^3}.
    \end{equation}
    Propagation of errors then gives us
    \begin{equation}
        \mu_2(h_{2,\mathrm{int}}) = \frac{(n-1)\mu_4 - (n-3)\mu_2^2}{n(n-1)}.
    \end{equation}
    In the right-hand side of this equation, $\mu_2$ and $\mu_4$ are the second and fourth true central moments of the observed velocity distribution.
    These are unknown, but we can approximate them with the $h$ statistics:
    \begin{equation}
        \mu_2(h_{2,\mathrm{int}}) \approx \frac{(n-1)h_4 - (n-3)h_2^2}{n(n-1)}.
    \end{equation}
    Even though the $h$ statistics are symmetrically unbiased estimators of the true central moments, the above approximation will have a bias because it is not a linear transformation.
    A further bias will be introduced by taking the square root to arrive at an estimate for the uncertainty:
    \begin{equation}
        \varepsilon(h_{2,\mathrm{int}}) \approx \sqrt{\mu_2(h_{2,\mathrm{int}})}.
    \end{equation}

    In a similar way, we can estimate the uncertainty on the fourth intrinsic moment.
    This calculation depends, however, on even higher moments, up to the eighth.
    With the small number of stars per bin in this paper, it is not feasible to calculate this uncertainty to a good accuracy.
    As the virial shape parameters depend on the fourth velocity moments, the uncertainty on the virial shape parameters will also be challenging to constrain.
    We therefore opt not to use the virial shape parameters in this paper.

    Calculating the above estimators and uncertainty on mock data drawn from known generalized normal distributions representing the intrinsic velocity distributions and normal distributions representing the measurement uncertainties, both similar to the properties of the observed data, shows that the intrinsic moments and the uncertainty on the second moment can on average be recovered with at most a few percent bias, which is much smaller than the statistical uncertainties.
    However, it is possible that by the subtraction of the cumulant correction the estimated moments become negative, whereas from equation \eqref{eq:infmom} it is clear that even moments of real-valued distributions must be non-negative.
    This is unavoidable when the statistical uncertainty of a moment is similar to or larger than the moment itself.

\section{Supplementary figures of CJAM parameter constraints}
\label{app:cjamsuppl}
\begin{figure}
        \includegraphics[width=\linewidth]{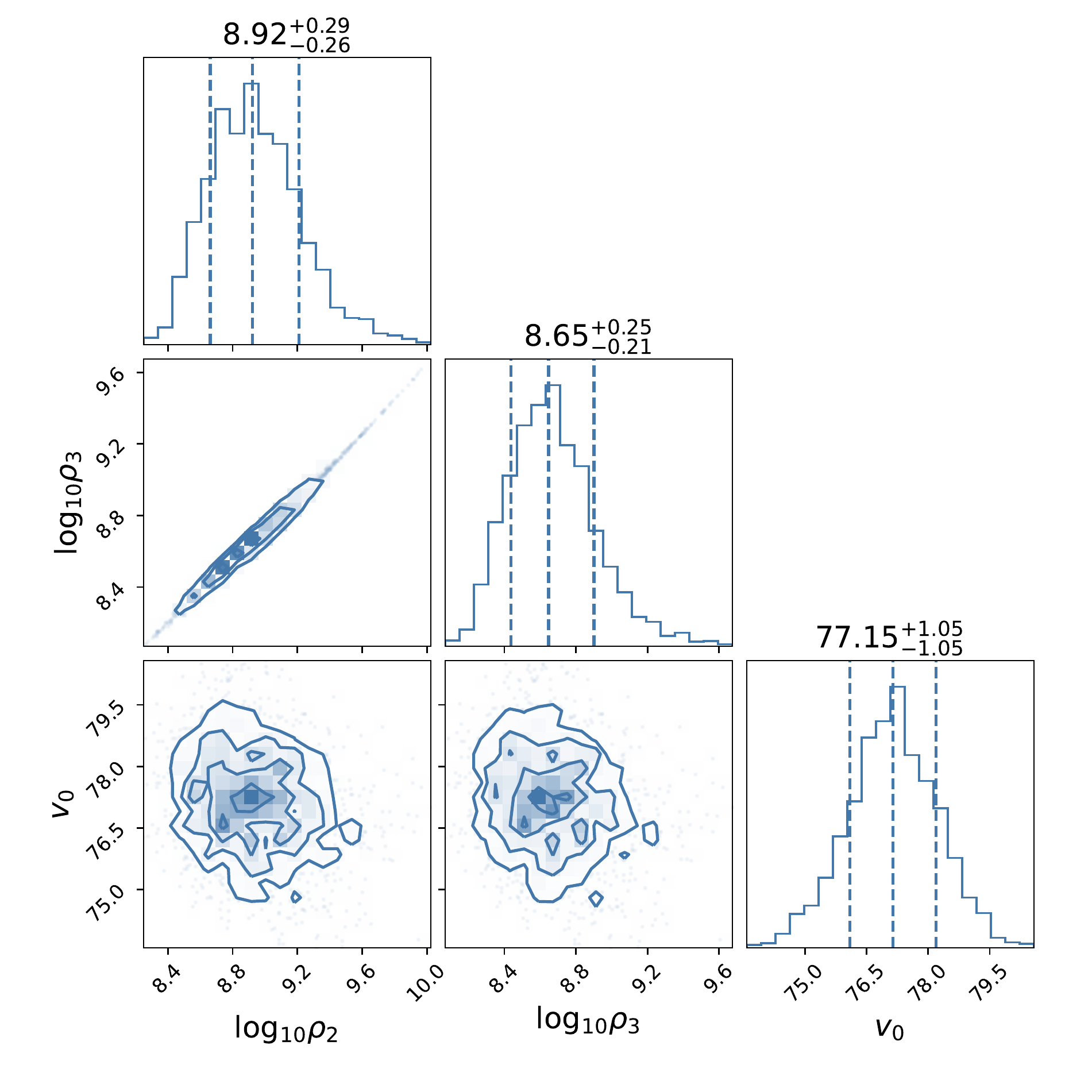}%
        \caption{%
            Constraints on the dark matter--density profile of Eridanus~2 in the computational parametrization, assuming cold dark matter, found using CJAM and MultiNest.
            Units are omitted for clarity.
            The parameters are the dark-matter density~$\rho_2$ and~$\rho_3$ respectively at~$100\,\mathrm{pc}$ and~$150\,\mathrm{pc}$ in $M_\odot\,\mathrm{kpc}^{-3}$ and the systemic velocity~$v_0$ in $\mathrm{km}\,\mathrm{s}^{-1}$.
            The contours correspond to $0.5\sigma$, $1.0\sigma$, $1.5\sigma$, and $2.0\sigma$ confidence levels, where $\sigma$ is the standard deviation of a two-dimensional normal distribution.
            The vertical dashed lines in the panels on the diagonal indicate the median and 68-\% confidence interval.%
        }
        \label{fig:cdmcornersupl}
    \end{figure}

\begin{figure*}
        \includegraphics[width=0.5\linewidth]{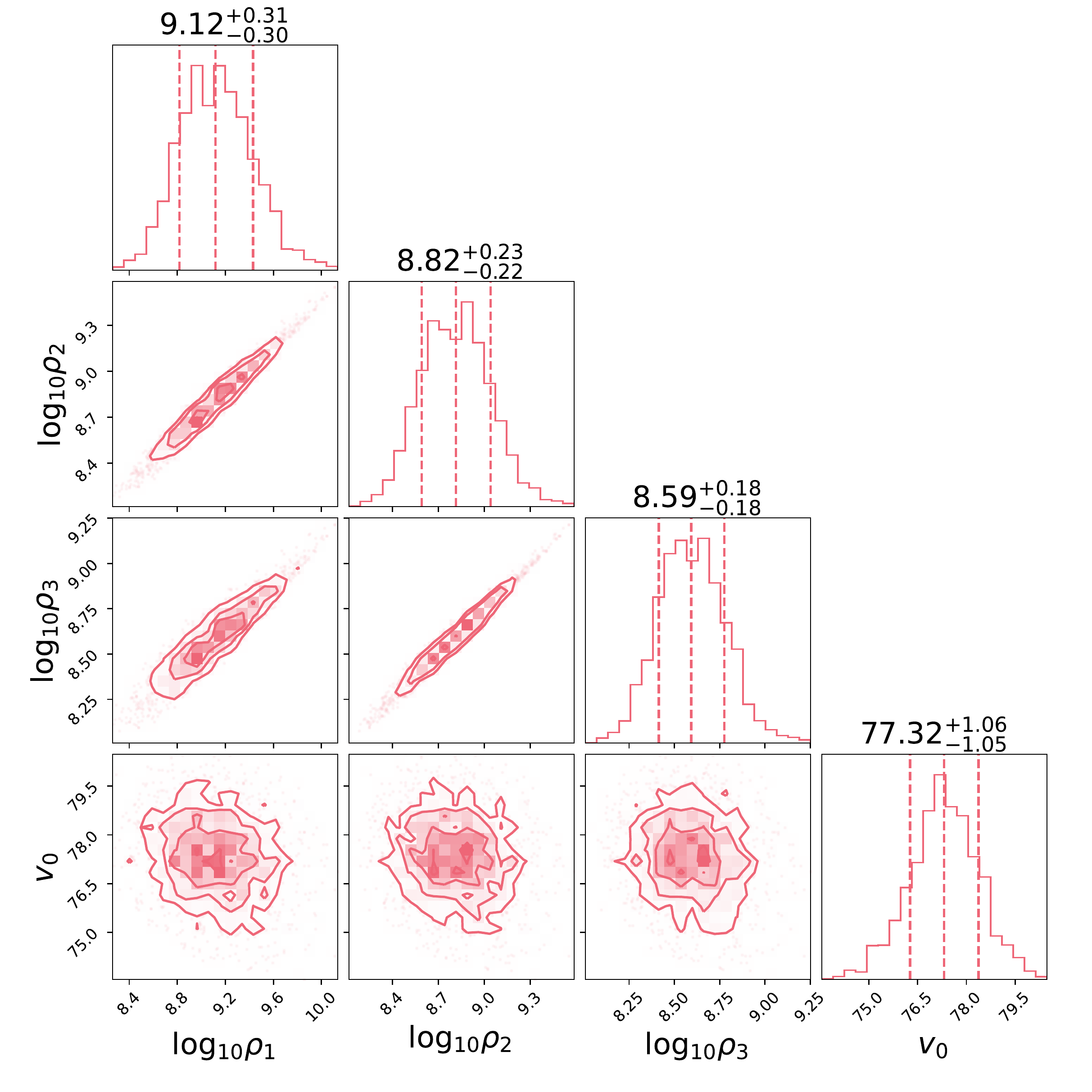}%
        \includegraphics[width=0.5\linewidth]{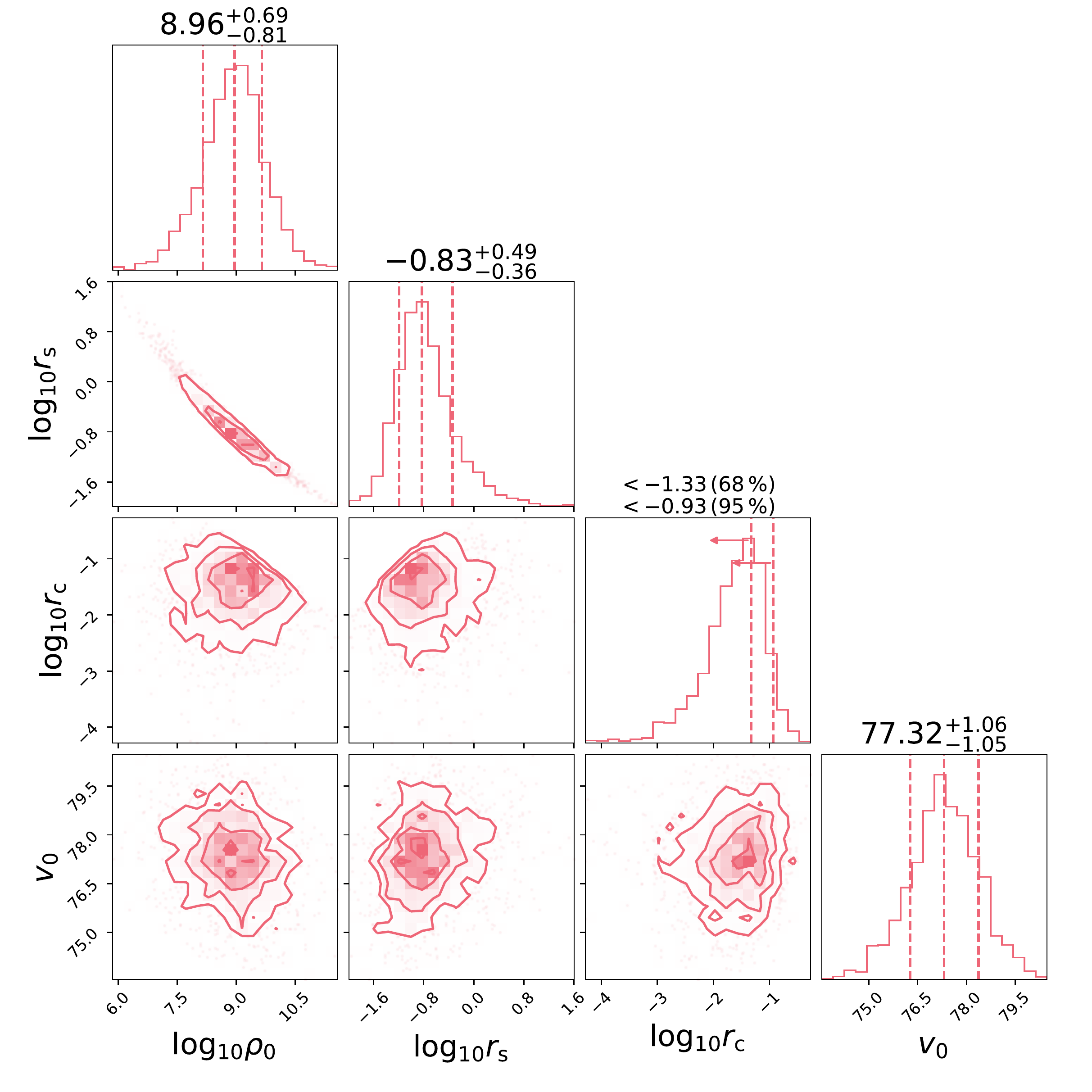}%
        \caption{%
            Constraints on the dark matter--density profile of Eridanus~2 in the computational (\textit{left}) and astrophysical (\textit{right}) parametrizations, assuming self-interacting dark matter, found using CJAM and MultiNest.
            Units are omitted for clarity.
            The parameters are the dark-matter density~$\rho_1$, $\rho_2$, and $\rho_3$ respectively at~$50\,\mathrm{pc}$, $100\,\mathrm{pc}$, and $150\,\mathrm{pc}$ in $M_\odot\,\mathrm{kpc}^{-3}$, the characteristic dark-matter density~$\rho_0$ in $M_\odot\,\mathrm{kpc}^{-3}$, the scale radius~$r_\mathrm{s}$ and core radius~$r_\mathrm{c}$ in $\mathrm{kpc}$, and the systemic velocity~$v_0$ in $\mathrm{km}\,\mathrm{s}^{-1}$.
            The contours correspond to $0.5\sigma$, $1.0\sigma$, $1.5\sigma$, and $2.0\sigma$ confidence levels, where $\sigma$ is the standard deviation of a two-dimensional normal distribution.
            The vertical dashed lines in the panels on the diagonal indicate the median and 68-\% confidence interval (without arrows) or the 68-\% and 95-\% confidence limits (upper and lower arrows, respectively).%
        }
        \label{fig:sidmcornersupl}
    \end{figure*}

\begin{figure*}
        \includegraphics[width=0.5\linewidth]{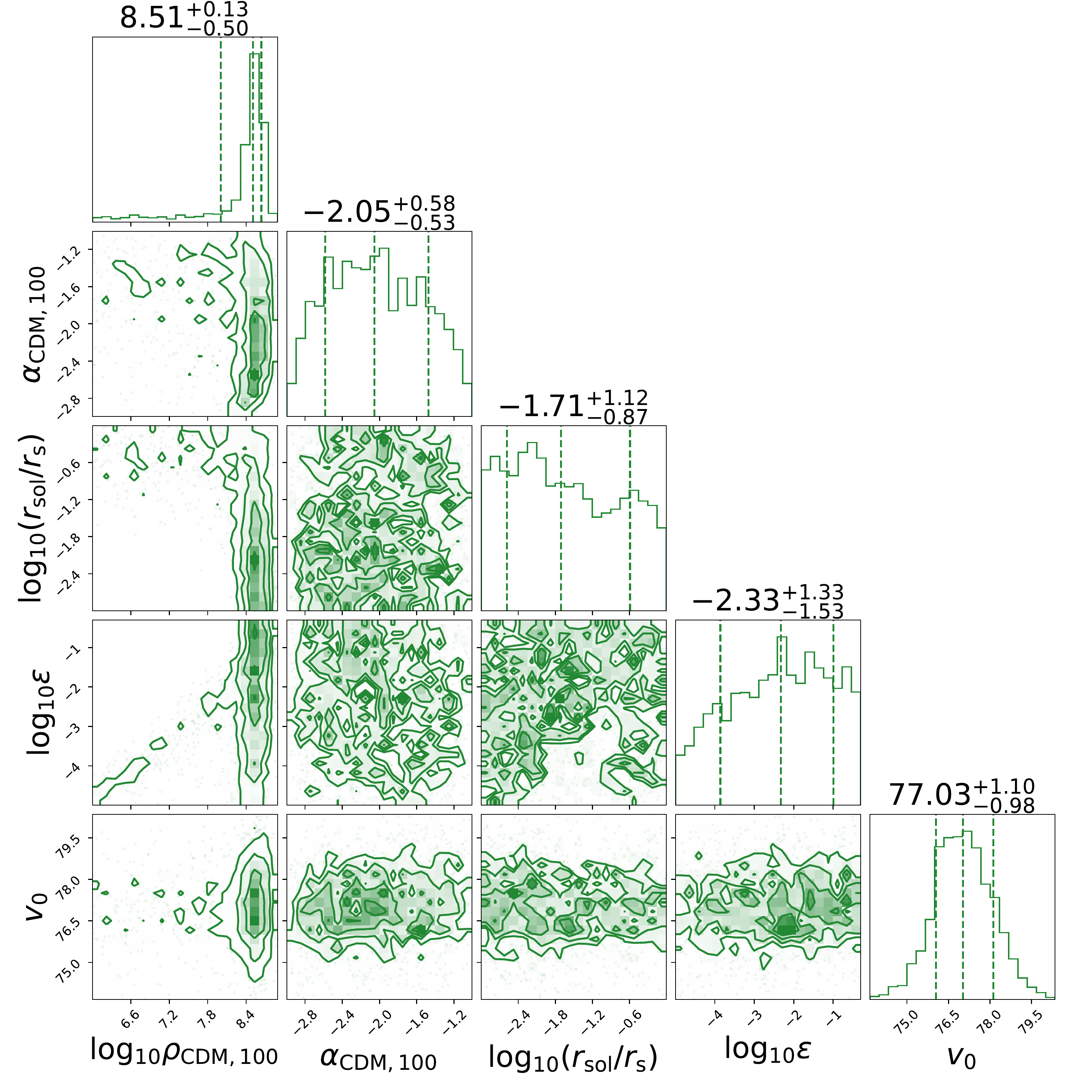}%
        \includegraphics[width=0.5\linewidth]{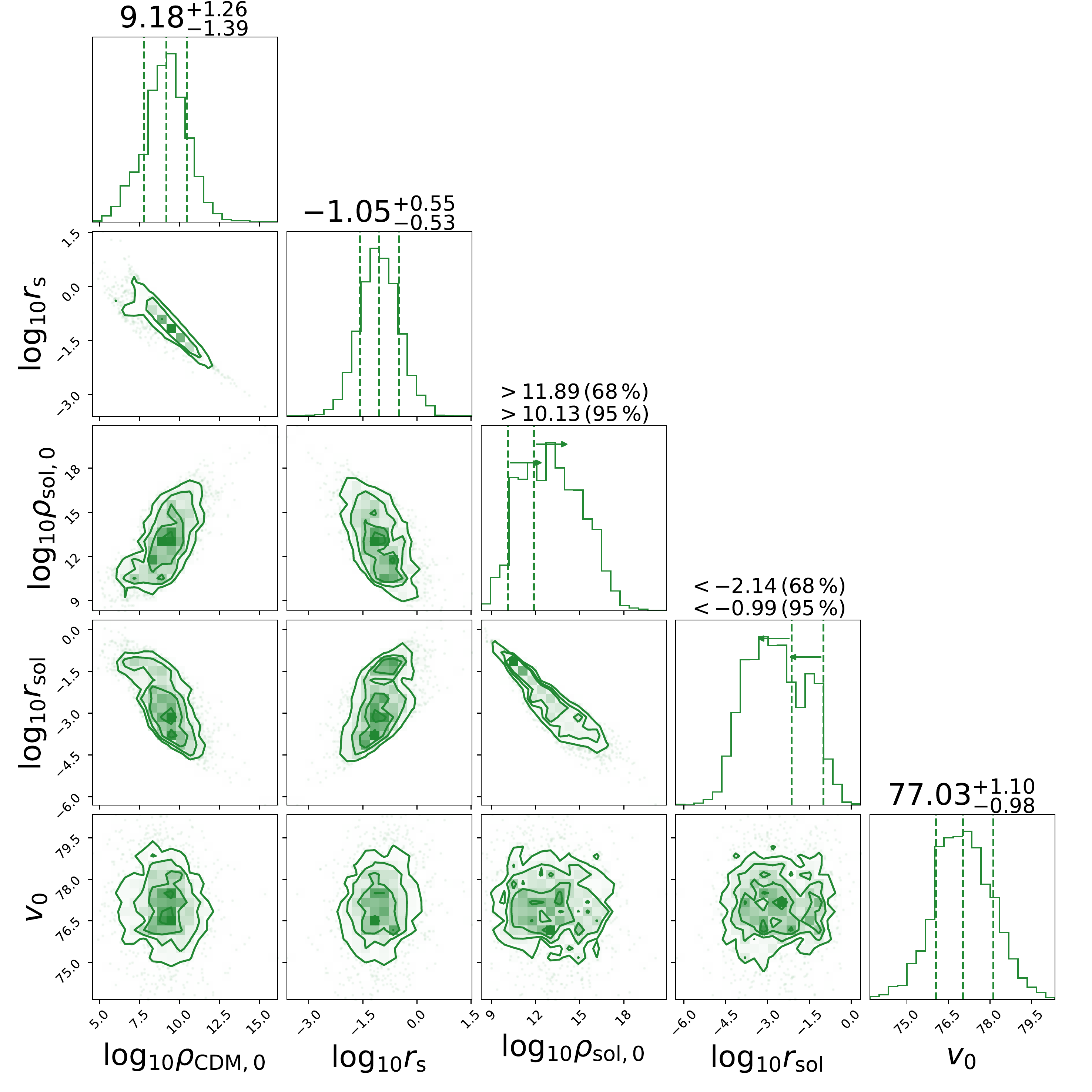}%
        \caption{%
            Constraints on the dark matter--density profile of Eridanus~2 in the computational (\textit{left}) and astrophysical (\textit{right}) parametrizations, assuming fuzzy dark matter, found using CJAM and MultiNest.
            Units are omitted for clarity.
            The parameters are the dark-matter density~$\rho_{\mathrm{CDM},100}$ of the CDM-like outer profile at~$100\,\mathrm{pc}$, the logarithmic slope~$\alpha_{\mathrm{CDM},100}$ of the CDM-like outer profile at $100\,\mathrm{pc}$, the soliton radius~$r_\mathrm{sol}$ in $\mathrm{kpc}$, the scale radius~$r_\mathrm{s}$ of the CDM-like outer profile in $\mathrm{kpc}$, the ratio $\varepsilon$ of the dark-matter density at the transition between inner and outer profile over the central soliton density, the characteristic dark-matter density~$\rho_{\mathrm{CDM},0}$ of the CDM-like outer profile in $M_\odot,\mathrm{kpc}^{-3}$, the central dark-matter density~$\rho_{\mathrm{sol},0}$ of the soliton in $M_\odot\,\mathrm{kpc}^{-3}$, and the systemic velocity~$v_0$ in $\mathrm{km}\,\mathrm{s}^{-1}$.
            The contours correspond to $0.5\sigma$, $1.0\sigma$, $1.5\sigma$, and $2.0\sigma$ confidence levels, where $\sigma$ is the standard deviation of a two-dimensional normal distribution.
            The vertical dashed lines in the panels on the diagonal indicate the median and 68-\% confidence interval (without arrows) or the 68-\% and 95-\% confidence limits (upper and lower arrows, respectively).%
        }
        \label{fig:fdm4cornersupl}
    \end{figure*}

    In this appendix we show the CJAM constraints on the dark matter--density profile of Eri~2 in additional parametrizations.
    Fig.~\ref{fig:cdmcornersupl} shows constraints for the CDM model in the computational parametrization.
    Figs.~\ref{fig:sidmcornersupl} and~\ref{fig:fdm4cornersupl} show constraints for SIDM and FDM, respectively, in the computational and astrophysical parametrizations.

\section{Recovery of intrinsic velocity dispersion profiles}
\label{app:vdisp}
\begin{figure}
        \includegraphics[width=\linewidth]{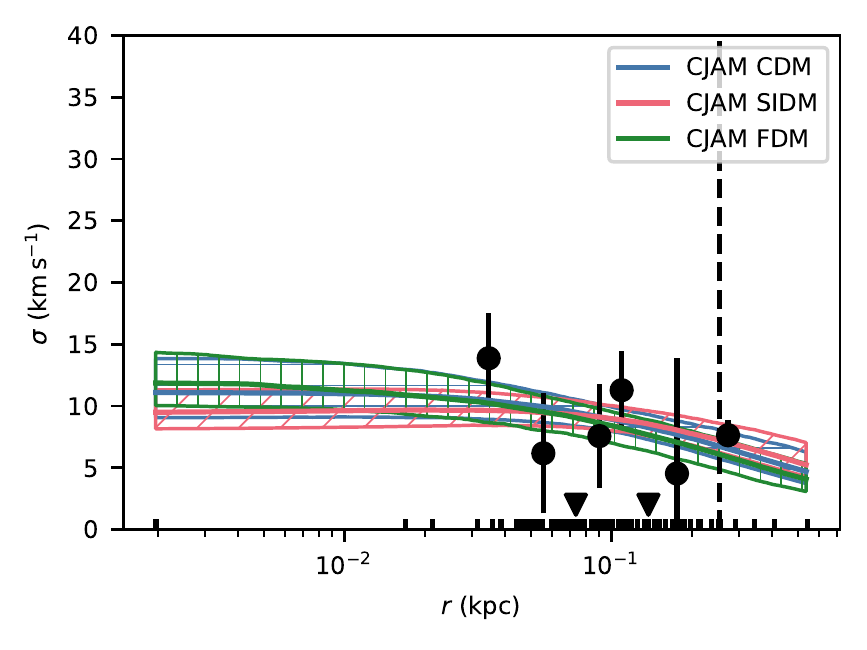}
        \includegraphics[width=\linewidth]{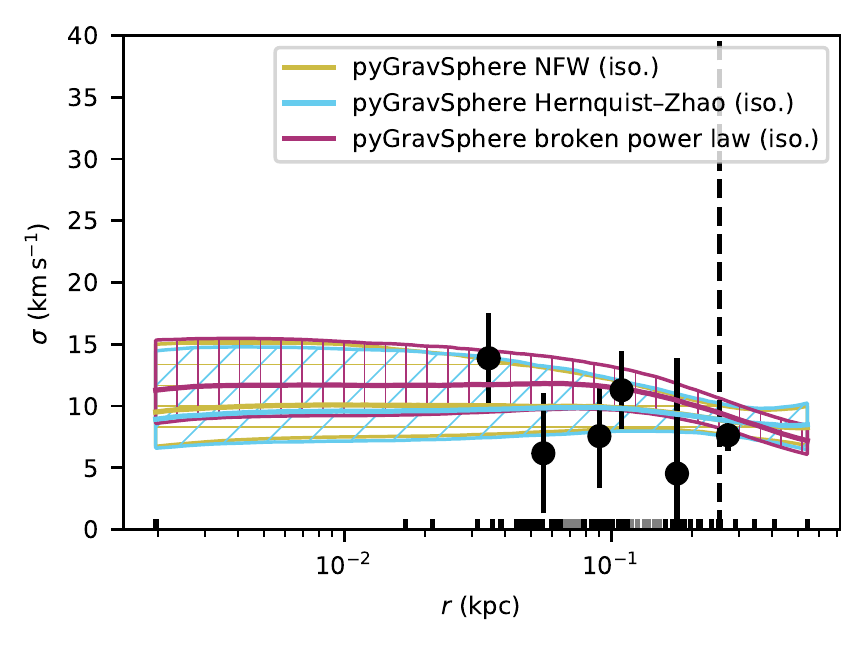}
        \includegraphics[width=\linewidth]{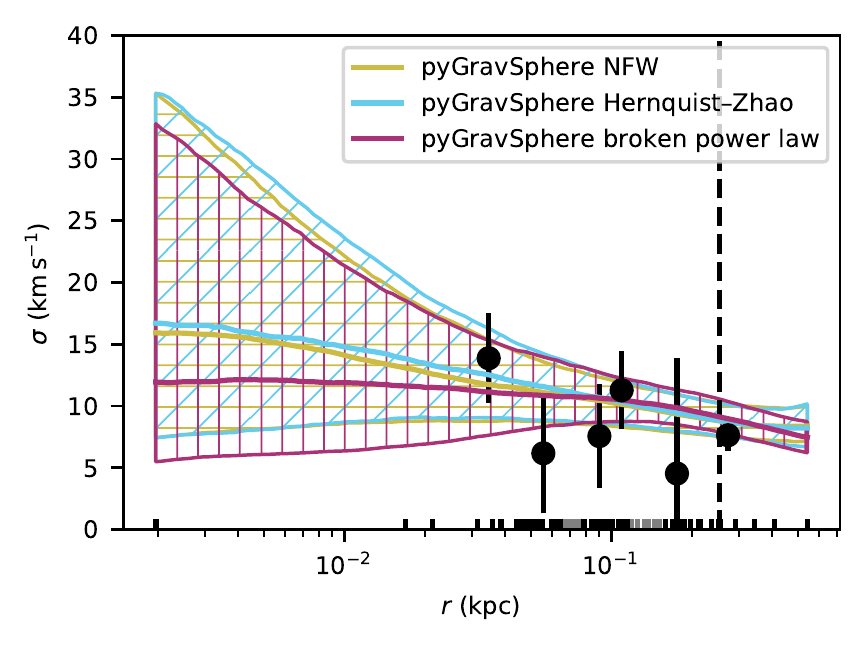}%
        \caption{%
            Recovered intrinsic velocity dispersion profiles of Eridanus 2.
            (\textit{top}) CJAM models for cold dark matter~(CDM), self-interacting dark matter~(SIDM), and fuzzy dark matter~(FDM).
            (\textit{centre}) pyGravSphere models assuming an isotropic velocity distribution, with Navarro--Frenk--White~(NFW), Hernquist--Zhao, and broken power-law profiles.
            (\textit{bottom}) As above, without assuming isotropy.
            Binned intrinsic velocity dispersion estimates are indicated with black circles and error bars and downward triangles where negative.
            The hatched bands represent the 68-\% confidence interval on the density at each radius.
            The half-light radius is indicated with the vertical dashed line.
            The black markers at the bottom of the figure show the projected radii of the kinematic tracers.
            Tracers in bins rejected by pyGravSphere are marked in grey.%
        }
        \label{fig:vdisp}
    \end{figure}

    Profiles of the intrinsic velocity dispersion allow for direct comparison between models and data-derived estimates.
    In Fig.~\ref{fig:vdisp} we display this comparison.
    In addition to the CJAM models with isotropic velocities and the pyGravSphere models with anisotropic velocities, used in the main body of this paper, we also display pyGravSphere models with isotropic velocities for comparison.
    The assumption on the velocity distribution has a large effect on the uncertainty in the intrinsic velocity dispersion at small radii, but the profiles are in all cases consistent with each other within their uncertainties.
    To compare the recovered profiles to the measured data, we show the estimated intrinsic velocity dispersion and its uncertainty in each pyGravSphere bin.
    For pyGravSphere we do not display bins with negative estimates (which are unphysical).
    We remind the reader that CJAM does not bin the velocity data; CJAM does not directly fit to the estimates displayed here.
    For CJAM we indicate the negative estimates as well.

    The intrinsic velocity dispersion profiles clarify the origin of some of the differences in the density profiles.
    The difference in the scale radius, with CJAM preferring smaller values than pyGravSphere, seems to be driven by the outer bins.
    The unbinned analysis of CJAM recovers an intrinsic dispersion profile that is lower at large radii, while pyGravSphere prefers models that are flatter.
    The higher density of the broken power-law profile around $100\,\mathrm{pc}$ is also visible in the dispersion profile and seems to be the result of overfitting to the estimators.
\end{document}